# MESOSCOPIC STRUCTURE OF MIXED TYPE DOMAIN WALLS IN MULTIAXIAL FERROELECTRICS

Anna N. Morozovska [1*], Eugene A. Eliseev[2†], Yevhen M. Fomichov[3], and Sergei V. Kalinin[4‡]

[1] Institute of Physics, National Academy of Sciences of Ukraine,

46, pr. Nauky, 03028 Kyiv, Ukraine

[2] Institute for Problems of Materials Science, National Academy of Sciences of Ukraine,

Krjijanovskogo 3, 03142 Kyiv, Ukraine

[3] Charles University in Prague, Faculty of Mathematics and Physics,

V Holešovičkach 2, Prague 8, 180 00, Czech Republic

[4] The Center for Nanophase Materials Sciences, Oak Ridge National Laboratory,

Oak Ridge, TN 37922

## Abstract

The structure of a 180-degree uncharged rotational domain wall in a multiaxial ferroelectric film is studied in the framework of an analytical Landau-Ginzburg-Devonshire (LGD) approach. Finite Element Modelling (FEM) is used to solve numerically the system of the coupled nonlinear Euler-Lagrange (EL) second order differential equations for two components of polarization. We show that the structure of the domain wall and corresponding metastable or stable phase of the film are controlled by a single parameter, − the dimensionless ferroelectric anisotropy factor $\mu$. We fitted the static profile of a solitary domain wall, calculated by FEM, with kink-like functions for polarization components, and extracted the five $\mu$-dependent parameters from the fitting to FEM curves. The surprisingly high accuracy of the fitting results for two polarization components in the entire $\mu$-range allows us to conclude that the analytical functions, which are trial functions in the direct variational method, can be treated as a high-accuracy variational solution of the static EL equations. We further derive exact two-component analytical solutions of the static EL equations for a polydomain 180-degree domain structure in a multiaxial ferroelectric film. Using these, we derive analytical expressions for the system free energy and analyze its dependence on the film thickness and boundary conditions at the film surfaces. The single-domain state is ground for zero polarization derivative at the surfaces, while the polydomain states minimize the system energy for zero polarization at the surfaces. Counterintuitively, the energy of the polydomain states split into two levels "0" and "1" for zero polarization at the surfaces, and each of the levels contains a large number of close-energy sub-levels, which structure is characterized by different number and type of the domain walls. The analytical solutions can become a useful tool for Bayesian analysis of HR STEM images in ferroelectric films.

---

\*      Corresponding author 1: anna.n.morozovska@gmail.com
†      Corresponding author 2: eugene.a.eliseev@gmail.com
‡      Corresponding author 3: sergei2@ornl.gov

# 1. INTRODUCTION

Multiaxial ferroelectrics are one of the most fascinating representatives of materials with multiple interacting order parameters and key objects for fundamental exploration of nonlinear and cooperative phenomena at micro, nano and atomic scales [1, 2, 3]. These materials undergo a temperature-driven phase transition accompanied by the appearance of a spontaneous polarization vector [4, 5]. The spatial distribution of the spontaneous polarization is often characterized by a complex morphology of domain structure [6, 7, 8, 9] and its nontrivial temporal evolution [10, 11]. This interest is further stimulated by multiple applications in extant and emerging technologies in industrial, medical and consumer sectors, including transducers, filters, sensors, ultrasonic motors and actuators, electronics and information storage [12, 13].

Theoretical modeling and practical control of the domain structure in multiaxial ferroelectrics is interesting from the fundamental viewpoint and also is important for many applications [4, 5, 14, 15, 16, 17, 18, 19], but it is strongly complicated due to a wide range of contributing physical processes. These involve interaction of domain walls with lattice potential barriers [20], point and planar defects [21, 22, 23] including charged acceptor/donor impurities and vacancies [24, 25], electric and elastic dipolar defects [26, 27, 28], twin and grain boundaries [29], as well as screening conditions at surfaces and interfaces [30, 31]. In ferroelectric thin films and their multilayers, which are intriguing objects of fundamental research and promising materials for nanoelectronics, decreasing the thickness usually leads to the ferroelectricity suppression and critically influences on the domain structure dynamics [32, 33, 34].

Seminal theoretical studies devoted to the structure of mixed **Ising-Bloch-type** [35, 36] and **Ising-Bloch-Neel-type** [37, 38] rotational domain walls in multiaxial ferroelectrics have been performed using the continuum Landau-Ginzburg-Devonshire (**LGD**) theory implemented to the Finite Element Modelling (**FEM**) or phase-field algorithms [17]. The approach allows obtaining accurate numerical results. However, due to the very complex nature of the abovementioned phenomena, the analytical theory of domain structure thermodynamics and kinetics is in multiaxial ferroelectrics thin films is studied only weakly, with exceptions for several special cases [39, 40, 41]. Partially this dearth of studies has been related to the lack, until recently, of direct information on domain wall structure.

Recently, the emergence of the high-resolution scanning transmission electron microscopy (**HR STEM**) has allowed direct insight into the structure of ferroelectric domain walls and interfaces at the atomic level. Enabled by the pioneering work by Jia et al. [42, 43, 44], Chisholm et al. [45], Nelson et al. [46, 47], Das et al. [48], and others [49, 50, 51] over decade ago, the direct studies of domain wall structures have become common. It has been shown that the comparison of the experimentally observed domain wall profiles with the functional forms derived from analytical theory allows determination of the Landau-Ginzburg functional parameters, such as gradient terms and boundary conditions at the interfaces [52, 53]. Very recently, this approach was extended towards the Bayesian context and the conditions for reliable elucidation of the unknown physics of materials system [54, 55]. However, the availability of high-quality experimental data necessitates further development of high-veracity universal analytical models for domain wall structure.

To fill the gap in the knowledge, here we consider the dynamics of a 180-degree uncharged rotational domain wall in a miltiaxial ferroelectric film within the framework of analytical LGD approach (see **Section 2**). FEM is used to solve numerically the system of the coupled nonlinear Euler-Lagrange (**EL**) differential equations of the second order for two components of polarization (see **Section 3**). Next, using LGD approach, we derived and analyzed the analytical solutions of the static EL equations for a polydomain 180-degree domain structure in a multiaxial ferroelectric film, which contain enough free parameters to satisfy arbitrary boundary conditions at the film surfaces (see **Section 4**). We analyze the domain state free energy dependence on the film thickness and boundary conditions. **Section 5** illustrates how the analytical solutions can be used for a Bayesian analysis of HR STEM data in thin ferroelectric films. **Section 6** is a brief summary. Calculation details of analytical solutions and free energy with renormalized coefficients obtained by direct variation method are listed in **Appendixes A** and **B** in **Suppl. Materials**, respectively.

## 2. PROBLEM FORMULATION

Here we consider a stress-free ferroelectric film (or layer) under the absence of free charges in it, and assume that the polarization vector **P** of the ferroelectric has a specific structure leading to the absence of a depolarization field, namely $div\mathbf{P} = 0$. The assumption corresponds to the case of uncharged domain walls, and is consistent with the ideal screening at the interfaces. However, in general case the depolarization fields will be present. External field is absent, since we are interested in a domain structure relaxation to a static equilibrium picture.

Without electrostriction and flexoelectric coupling, the explicit expressions of LGD energy density for polarization **P** has the form:

$$G = \int(g_L + g_{grad})dv + \int g_S ds. \qquad (1a)$$

Expressions for Landau ($g_L$), gradient ($g_{grad}$) and surface ($g_S$) energies in the ferroelectrics with the second order paraelectric-ferroelectric phase transition are

$$g_L = a_i P_i^2 + a_{ij} P_i^2 P_j^2 - P_i E_i, \qquad g_{grad} = \frac{g_{ijkl}}{2}\frac{\partial P_i}{\partial x_j}\frac{\partial P_k}{\partial x_l}, \qquad g_S = \frac{a_i^S}{2}P_i^2, \qquad (1b)$$

where summation is performed over all repeated indexes. As a rule, the coefficients $a_i$ linearly depend on the temperature $T$, $a_i = a_{Ti}(T - T_C)$; other coefficients are temperature independent, but can be affected by elastic strains via electrostriction effect [4, 5]. The surface energy coefficients $a_i^S$ are prior unknown because they depend on the surface/interface chemistry.

Dynamic equations of state follow from the variation of action $S$

$$S = \int_0^\infty dt \int \left[ g_L + g_{grad} - \frac{\rho_i}{2} \left( \frac{\partial P_i}{\partial t} \right)^2 \right] dv. \qquad (2)$$

The coefficient $\rho_i > 0$ in the kinetic term, $\frac{\rho_i}{2} \left( \frac{\partial P_i}{\partial t} \right)^2$.

Below we consider the two-component and one-dimensional case, when the polarization vector $\boldsymbol{P}(x_3)$ depends only on the coordinate $x_3$ that is normal to the film surface (see **Fig. 1**). The vector structure $\boldsymbol{P}(x_3) = \{P_1(x_3), P_2(x_3), 0\}$ is consistent with the absence of depolarization field, since $div\boldsymbol{P} \equiv 0$ in the case.

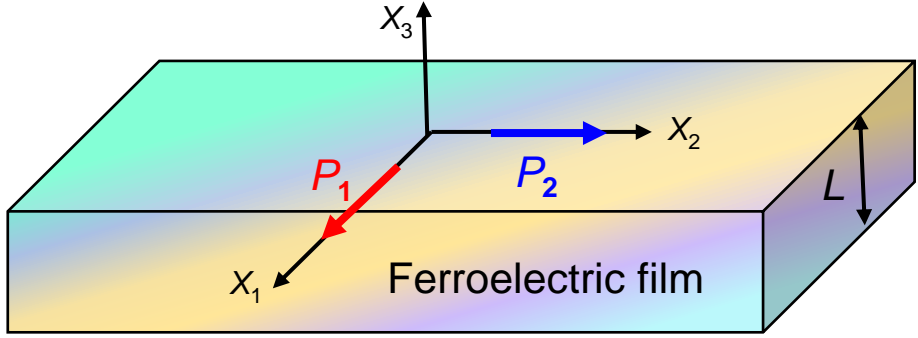

**FIGURE 1.** The geometry of a considered multiaxial ferroelectric film with polarization vector $\boldsymbol{P}(x_3) = \{P_1(x_3), P_2(x_3), 0\}$. The film is regarded enough thick, namely its thickness $L \gg R_c$, where $R_c$ is a correlation length.

The coupled time-dependent EL equations obtained from the variation of the action (2), allowing for the Landau-Khalatnikov relaxation [56], acquire a relatively simple form:

$$2a_1 P_1 + 4a_{11} P_1^3 + 2a_{12} P_1 P_2^2 - g_{44} \frac{\partial^2 P_1}{\partial x_3^2} = -\rho_1 \frac{\partial^2 P_1}{\partial t^2} - \Gamma \frac{\partial P_1}{\partial t}, \qquad (3a)$$

$$2a_1 P_2 + 4a_{11} P_2^3 + 2a_{12} P_2 P_1^2 - g_{44} \frac{\partial^2 P_2}{\partial x_3^2} = -\rho_1 \frac{\partial^2 P_2}{\partial t^2} - \Gamma \frac{\partial P_2}{\partial t}, \qquad (3b)$$

where $\rho_1 > 0$ is the kinetic coefficient, and $\Gamma > 0$ is the Khalantikov coefficient. Below we assumed m3m cubic symmetry of the high temperature phase, i.e. $g_L = \sum_{i,j \neq i} (a_1 P_i^2 + a_{11} P_i^4 + a_{12} P_i^2 P_j^2 - P_i E_i)$ .

The nonlinear coupled equations (3) are supplemented by boundary conditions at the film surfaces $x_3 = \mp \frac{L}{2}$:

$$\left( a_1^S P_i \mp g_{44} \frac{\partial P_i}{\partial x_3} \right)\bigg|_{x_3 = \mp \frac{L}{2}} = 0, \qquad i = 1, 2. \qquad (4a)$$

We note that $a_{11} > 0$, $a_{12} > -2a_{11}$, $g_{44} > 0$, $g_{11} > 0$, and $a_1^S \geq 0$ for the system stability; and consider negative coefficients $a_1 < 0$ corresponding to the bulk ferroelectric state at $T < T_C$. The boundary conditions (4a) significantly influence domain structure and evolution only in the case of relatively thin films, namely when the film thickness $L$ is smaller than the several tens of correlation length $R_c = \sqrt{-g_{44}/(2a_1)}$, e.g. $L \leq 100 R_c$.

For thick films with $L \geq 10^3 R_c$ the conditions (4a) have very little influence, and should be substituted by the conditions of periodicity or anti-periodicity for every polarization component and its derivative:

$$P_i\left(-\frac{L}{2}\right) = \pm P_i\left(+\frac{L}{2}\right), \quad \frac{\partial}{\partial x_3} P_i\left(-\frac{L}{2}\right) = \mp \frac{\partial}{\partial x_3} P_i\left(+\frac{L}{2}\right), \; i = 1, 2. \qquad (4b)$$

Here the sign "+" corresponds to the periodic, and the sign "–" – to the antiperiodic boundary condition. The sign choice is individual for each component. Note that the conditions (4b) are not equivalent to the conditions (4a), because (4b) are parity/periodicity conditions only. Complementary to the conditions (4b) one often need to impose some additional conditions on the polarization components, consistent with initial conditions.

The first integral of the system (3) is $I(x) = g_L - g_{grad}$. The conservation of this dynamic invariant, $I(x) = const$, means that the system tendency to lower the domain wall energy (coming from the gradient energy $g_{grad}$) is compensated by the ferroelectric nonlinearity energy (coming from the Landau energy $g_L$). For FEM simulations, initial distributions are taken either in the form of a random small-amplitude polarization, or in the form of a solitary domain wall with superimposed random fields, or in the form of a small-amplitude sinusoidal modulation.

## 3. EQUILIBRIUM STRUCTURE OF TWO-COMPONENT DOMAIN WALLS

### A. Finite Element Modeling of a Solitary Domain Wall

Here we analyse the structure of the equilibrium uncharged 180-degree domain walls in the second order ferroelectric, which are relaxed solution of Eqs.(3). To analyse the domain wall structure, we introduce the dimensionless coordinate $x$, film thickness $l$, polarization components $p_1$ and $p_2$, ferroelectric anisotropy factor $\mu$, relaxation time $\tau$ and kinetic coefficient $\rho$:

$$x = \frac{x_3}{R_c}, \quad l = \frac{L}{R_c}, \quad p_1 = \frac{P_1}{P_S}, \quad p_2 = \frac{P_2}{P_S}, \quad \mu = \frac{a_{12}}{2a_{11}}, \quad \tau = -\frac{\Gamma}{a_1}, \quad \rho = -\frac{\rho_1}{a_1}. \tag{5}$$

Here $P_S = \sqrt{-a_1/(2a_{11})}$ is the spontaneous polarization value, and $R_c = \sqrt{-g_{44}/(2a_1)}$ is a correlation length [see **Appendix A** in Suppl. Mat. [57] for details]. For multiaxial perovskite-type ferroelectrics with ferroelectric Curie temperature > 400 K (e.g. BaTiO$_3$, (Pb,Zr)TiO$_3$, BiFeO$_3$, etc.), the values $P_S = (0.25 - 1)$C/m$^2$ and $R_c = (0.2 - 0.5)$nm at room temperature. Thus, the dimensionless thickness $l = 100$ corresponds to sufficiently thin films with $l = (20 - 50)$nm. The anisotropy factor $\mu$ vary from $-0.25$ to $+4.5$ and can be temperature-dependent.

The dimensionless EL equations, obtained from dynamic Eqs.(3), have the form:

$$-\rho \frac{\partial^2 p_1}{\partial t^2} - \tau \frac{\partial p_1}{\partial t} = -\frac{\partial^2}{\partial x^2} p_1 - p_1 + p_1^3 + \mu p_1 p_2^2, \tag{6a}$$

$$-\rho \frac{\partial^2 p_2}{\partial t^2} - \tau \frac{\partial p_2}{\partial t} = -\frac{\partial^2}{\partial x^2} p_2 - p_2 + p_2^3 + \mu p_2 p_1^2. \tag{6b}$$

To study the polarization relaxation to a stable or metastable state, we set $\rho = 0$ and chose the calculation time $t_{max}$ much higher than the time $\tau$ of the polarization relation to an equilibrium state, e.g. $t_{max} \gg 100\,\tau$. Initial distribution of polarization used in FEM is chosen in the form of a solitary Ising-type domain wall perturbed by a small fluctuation:

$$p_1(x, t = 0) = p_0 \tanh\left(\frac{x}{b}\right) + \delta p_1(x), \qquad p_2(x, t = 0) = \delta p_2(x), \tag{7a}$$

where the random fluctuation $|\delta p_{1,2}(x)| \ll |p_0|$

The boundary conditions (4a) in the dimensionless variables acquire the form:

$$\left(\frac{p_1}{\lambda_1} \mp \frac{\partial p_1}{\partial x}\right)\bigg|_{x=\mp\frac{l}{2}} = 0, \qquad \left(\frac{p_2}{\lambda_2} \mp \frac{\partial p_2}{\partial x}\right)\bigg|_{x=\mp\frac{l}{2}} = 0. \tag{7b}$$

Here $\frac{1}{\lambda_i} = \frac{a_i^S}{g_{44} R_c}$ ($i = 1, 2$) are the dimensionless inverse extrapolation lengths [58], which are not negative, and can vary in a very wide range, $0 \leq \lambda_i < \infty$, due to uncertainty of available experimental parameters. For the case $\frac{1}{\lambda_i} > 0$ and $l < 100$, either a solitary domain wall, or a periodic domain structure, appears after the polarization relaxation to an equilibrium state. In this work we mostly consider the case $\frac{1}{\lambda_i} = 0$ (i.e. very high $\lambda_i \to \infty$), which corresponds to the "natural" boundary conditions, $\frac{dp_i}{dx}\big|_{x=\pm l/2} = 0$ [i.e. zero surface energy in Eq.(1a)], and the alternative, $\lambda_i = 0$, which corresponds to $p_i|_{x=\pm l/2} = 0$.

The antiperiodic-periodic boundary conditions (4b) consistent with the initial condition (7a), which will be used for thick films with $l \gg 100$, have the form:

$$p_1\left(-\frac{l}{2}\right) = -p_1\left(\frac{l}{2}\right), \quad \frac{\partial p_1}{\partial x}\bigg|_{-\frac{l}{2}} = \frac{\partial p_1}{\partial x}\bigg|_{\frac{l}{2}}, \quad p_2\left(-\frac{l}{2}\right) = p_2\left(\frac{l}{2}\right), \quad \frac{\partial p_2}{\partial x}\bigg|_{-\frac{l}{2}} = -\frac{\partial p_2}{\partial x}\bigg|_{\frac{l}{2}}. \tag{7c}$$

The boundary problem (6)-(7) depends on the only control parameter – ferroelectric anisotropy factor $\mu$. The inequality $-1 < \mu$ should be valid for the system stability. The numerical solutions of Eqs.(6) are shown in **Fig.2a-f** for several values of the anisotropy factor $\mu$.

The dimensionless LGD free energy density and the first integral corresponding to Eqs.(6) have the form:

$$G = \int_{-l/2}^{l/2} g_V(x)dx + \frac{p_1^2(-l/2)}{\lambda_1} + \frac{p_2^2(-l/2)}{\lambda_2} + \frac{p_1^2(l/2)}{\lambda_1} + \frac{p_2^2(l/2)}{\lambda_2}, \quad (8a)$$

$$g_V = -\frac{1}{2}(p_1^2 + p_2^2) + \frac{1}{4}(p_1^4 + p_2^4) + \frac{\mu}{2}p_1^2 p_2^2 + \frac{1}{2}\left[\left(\frac{dp_1}{dx}\right)^2 + \left(\frac{dp_2}{dx}\right)^2\right], \quad (8b)$$

$$I_1[\mu] = -\frac{1}{2}(p_1^2 + p_2^2) + \frac{1}{4}(p_1^4 + p_2^4) + \frac{\mu}{2}p_1^2 p_2^2 - \frac{1}{2}\left[\left(\frac{dp_1}{dx}\right)^2 + \left(\frac{dp_2}{dx}\right)^2\right] = \begin{cases} \frac{-1}{2(1+\mu)}, & -1 < \mu < 1, \\ -\frac{1}{4}, & \mu > 1. \end{cases}$$

(8c)

The free energy (8) as a function of polarization components $p_1$ and $p_2$ is shown in **Fig. 2** for different $\mu$ values and zero gradients consistent with the case $\lambda_i \to \infty$. Two spatially-homogeneous phases (8) exist, namely:

1) Orthorhombic **O-phase** with the minimal energy density $g_{LGD} = -\frac{1}{2(1+\mu)}$ corresponding to polarization components $p_1 = p_2 = \pm\frac{1}{\sqrt{1+\mu}}$ (see **Figs. 2a-c**). O-phase is stable at $-1 < \mu < 1$.

2) Tetragonal **T-phase** with the minimal and energy density $g_{LGD} = -\frac{1}{4}$ corresponding to polarization components $p_1^2 = 1$, $p_2^2 = 0$, or $p_1^2 = 0$, $p_2^2 = 1$ (see **Figs. 2e-f**). T-phase is stable at $\mu > 1$. The O→T transition takes place at $\mu = 1$, when the 4 potential minima merges and transforms in a circle (see **Figs. 2d**).

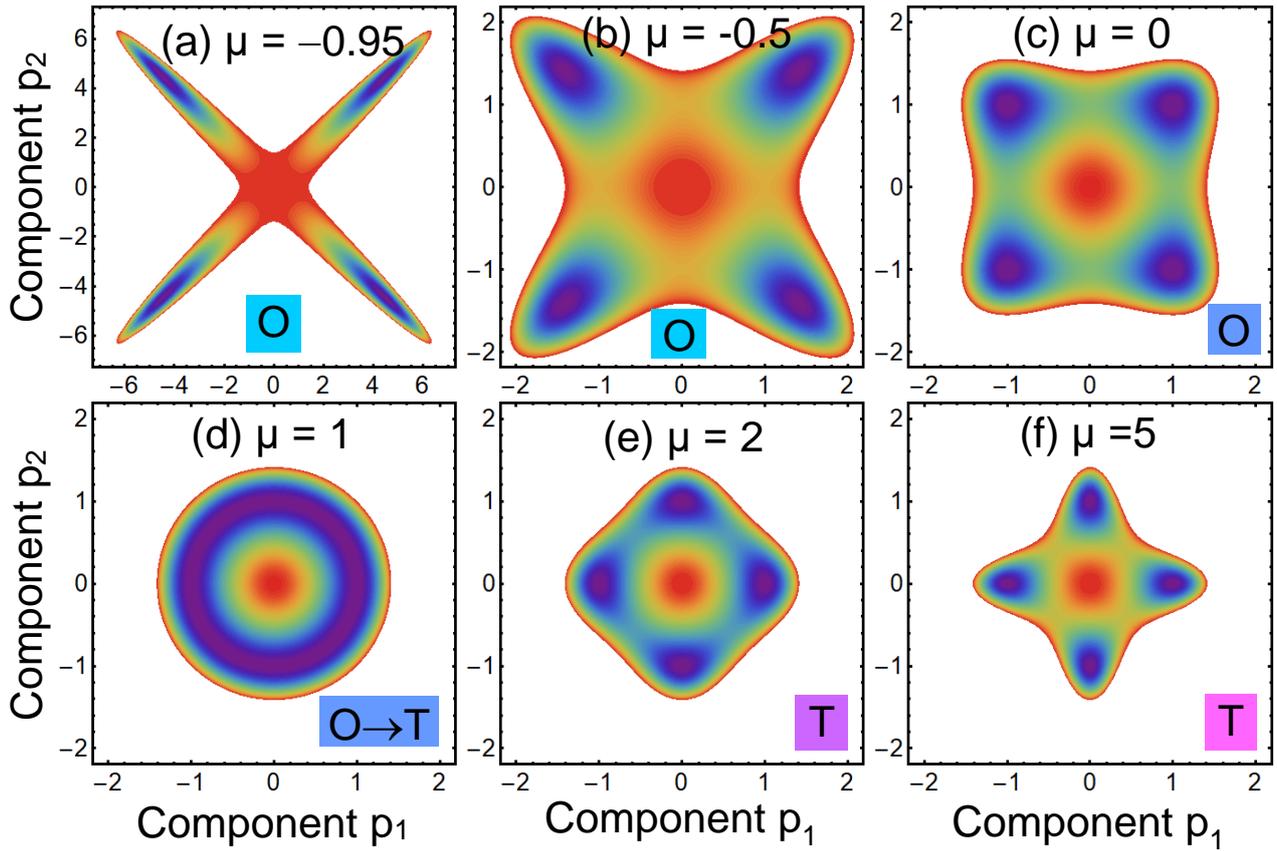

**FIGURE 2.** The free energy (8) as a function of order parameter components $p_1$ and $p_2$ for different values of parameter $\mu$: **(a)** $\mu = -0.95$, **(b)** $\mu = -0.5$, **(c)** $\mu = 0$, **(d)** $\mu = 1$, **(e)** $\mu = 2$ and **(f)** $\mu = 5$. Red color denotes zero energy, while violet color is its minimal density equal to $-10$ **(a)**, $-1$ **(b)**, $-1/2$ **(c)**, and $-1/4$ **(d, e, f)** relative units. Capital letters "O" and "T" denote orthorhombic and tetragonal spatially-homogeneous phases, respectively.

A detailed analysis of FEM results allows concluding that we can distinguish (somewhat arbitrarily) several different morphologies of the domain wall, shown in **Figs. 3a-f**, where the control parameter $\mu$ determines the structure of the uncharged 180-degree domain walls and values of polarization components $p_i$. The description of these areas is the following:

1) The first region "1", where $-1 < \mu < 0$, corresponds to the O-phase with Ising-Bloch domain walls (see **Fig. 3a**). Far from the wall (i.e. at $x \to \pm\infty$) the saturation expressions, $p_1 \to \pm\frac{1}{\sqrt{1+\mu}}$ and $p_2 \to +\frac{1}{\sqrt{1+\mu}}$, are valid for the polarization components. Thus $|p_1| = |p_2| > 1$ far from the wall. The component $p_1$ has an antisymmetric tanh-like profile across the domain wall; and the component $p_2$ has a symmetric profile with a sharp minimum well-localized at the wall. The minimum height decreases with $\mu$ increase, and disappears at $\mu \to 0$ (see **Fig. 3b**). Because $p_1$ is zero and $p_2$ is minimal at the wall, the contrast of the HR STEM image across the wall looks as "**dark-dark**" pattern for $-1 < \mu < 0$.

2) The second region "2", where $0 < \mu < 1$, corresponds to the O-phase with Ising-Bloch domain walls (see **Fig. 3c**). The expressions, $p_1 \to \pm\frac{1}{\sqrt{1+\mu}}$ and $p_2 \to \pm\frac{1}{\sqrt{1+\mu}}$, are valid far from the wall, where $|p_1| = |p_1| < 1$. The component $p_1$ has an antisymmetric tanh-like profile across the domain wall; and the component $p_2$ has a symmetric profile with a maximum at the wall. Both, the domain wall width for $p_1$ and $p_2$ components, and the maximum height for $p_1$ component, increase with $\mu$ increase; at that the wall becomes very thick and diffuse at $\mu \to 1$ (see **Fig. 3d**). At $\mu = 1$ the value $p_1^2 + p_2^2$ is invariant, and so the exceptional case of "isotropic" ferroelectric is realized by solution of Eqs.(6). Because $p_1$ is zero and $p_2$ has a sharp maximum at the wall, the contrast of the HR STEM image across the wall looks as "**dark-bright**" thin pattern for $0 < \mu < 1$.

3) The third region "3", where $1 < \mu < 3$, corresponds to the T-phase with mixed-type Ising-Bloch domain walls (see **Fig. 3e**). Far from the wall, where $p_1 \to \pm 1$ and $p_2 \to 0$. The component $p_1$ has an antisymmetric double-step-like profile across the domain wall; and the component $p_2$ has a symmetric profile with a flat maximum that is centered at the wall. Both the steps width for $p_1$, and the height and width of $p_2$ maximum, gradually decreases with $\mu$ increase, and eventually disappears at $\mu = 3$. An energy estimate shows that step-like Ising-Bloch and purely Ising walls coexist at $\mu = 3$. Because $p_1$ has a zero plateau and $p_2$ has a flat maximum at the wall, the contrast of the HR STEM image across the wall looks as "**dark-bright**" thick pattern for $1 < \mu < 3$.

4) The fourth region "4", where $\mu > 3$, corresponds to the T-phase with purely Ising-type domain walls (see **Fig. 3f**). The component $p_2$ is absent and the component $p_1$ has an antisymmetric tanh-like profile across the domain wall and saturates far from the wall, where $p_1 \to \pm 1$. Since $p_2 \equiv 0$, the domain wall profile is $\mu$-independent. Because $p_1$ is zero at the wall and $p_2$ is absent, the contrast of the HR STEM image across the wall looks as "**dark**" pattern for $\mu > 3$.

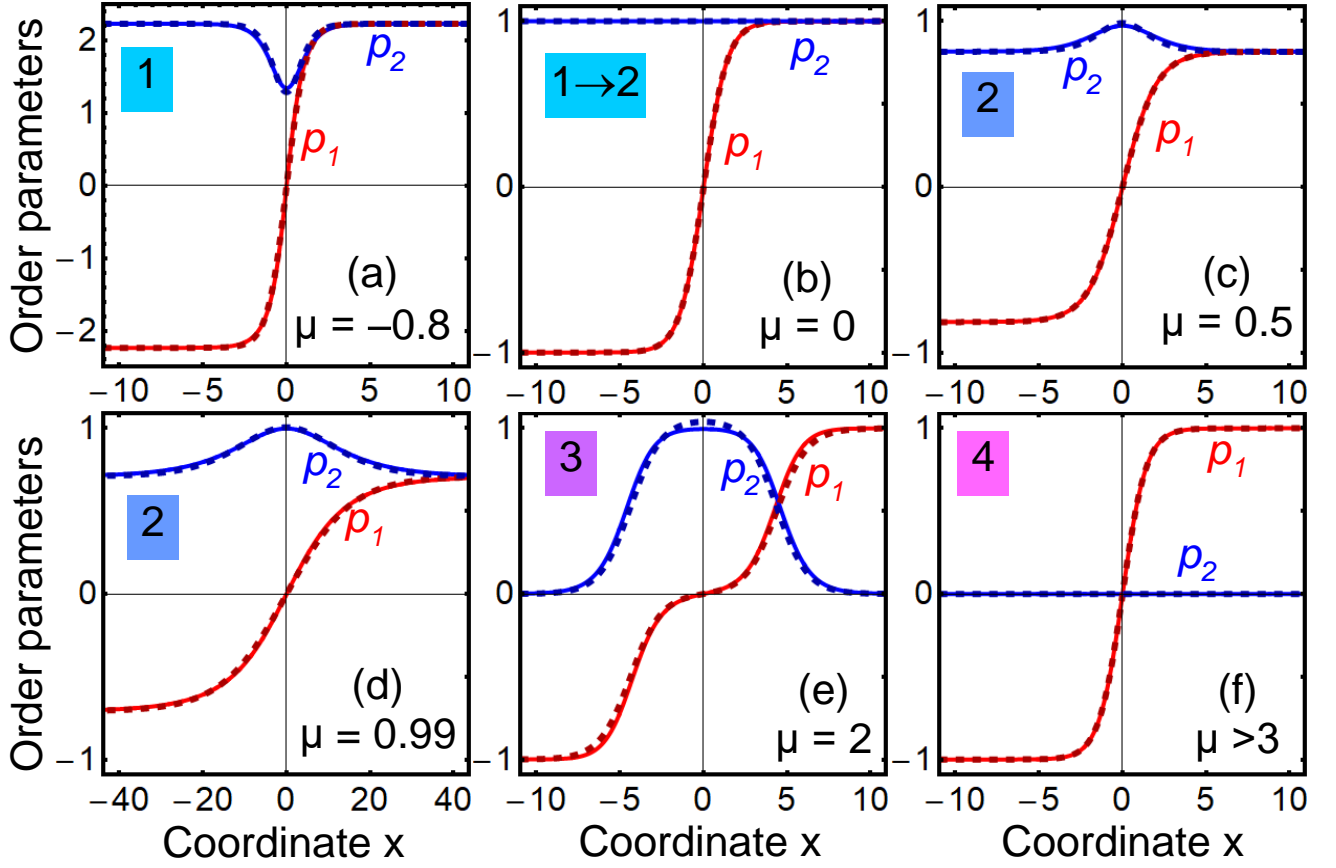

**FIGURE 3.** Profiles of polarization components $p_1$ (red curves) and $p_2$ (blue curves) calculated for different values of ferroelectric anisotropy $\mu$: **(a)** $\mu = -0.8$, **(b)** $\mu = 0$, **(c)** $\mu = 0.5$, **(d)** $\mu = 0.99$, **(e)** $\mu = 2$ and **(f)** $\mu = 4$. Solid and dashed curves represent FEM solution of Eqs.(6) and its fitting with the trial functions (9), respectively. Numbers 1 – 4 denote the regions with different morphology of domain wall: "1" is for the Ising-Bloch domain wall in the O-phase; "1→2" is for the Ising wall at the domain boundary; "2" is for the simple Ising-Bloch domain wall in the O-phase, "3" for is for the step-like Ising-Bloch-type domain wall in T-phase; and "4" is for the purely Ising domain wall in T-phase. Film thickness $l \gg 100$.

### B. Analytical Solutions for a Solitary Domain Wall in a Thick Film

To find equilibrium analytical solutions, direct variational method is applied for Eqs.(6) using the trial functions in the form of two kinks superposition and constants:

$$p_1(x) = a_0 + \frac{a_1}{2}\left[\tanh\left(\frac{x+x_w}{w}\right) + \tanh\left(\frac{x-x_w}{w}\right)\right], \quad (9a)$$

$$p_2(x) = a_2 + \frac{b_2}{2}\frac{w}{x_w}\left[\tanh\left(\frac{x+x_w}{w}\right) - \tanh\left(\frac{x-x_w}{w}\right)\right]. \quad (9b)$$

The constant amplitude $a_0 \equiv 0$ to satisfy the antisymmetric boundary conditions for $p_1(x)$ [see Eqs.(7c)]. The amplitudes $a_1$ and $a_2$ define polarization components far from the wall, because $p_1(x \to \pm\infty) \to \pm a_1$ and $p_2(x \to \pm\infty) \to a_2$. The amplitude $b_2$ contributes to the $p_2$ extremum at the wall, since $p_2(0) \to a_2 + b_2 \frac{w}{x_w} \tanh\left(\frac{x_w}{w}\right)$. The length $w$ and shift $x_w$ define the width of the $p_1(x)$ and $p_2(x)$ domain walls. These five variational parameters can be determined after substitution of Eqs.(9) in the free energy (7), further integration and minimization of the free energy over these parameters. This allows us to obtain analytical dependencies for the variational parameters on the control parameter $\mu$ (see **Appendix B** in Suppl. Mat.[57]).

The choice of the trial functions (9) is based on the fact that the functions are exact and stable solutions of the Eqs.(6) for zero anisotropy, $\mu = 0$, and relatively high anisotropy, $\mu \geq 3$. Corresponding values of parameters $a_1$, $a_2$, $b_2$, $w$, and $x_w$ are listed in **Table I.** The solution (9) also describes the instability limit at $\mu \to -1$ and the first order phase transition at $\mu = 1$ [see **Table I** for details]. When we use the antiperiodic-periodic boundary conditions (7c) for FEM, the shift $x_w$ should be zero for the stability of the numerical solution in thick films at $-1 < \mu < 1$ (O-phase) and $\mu > 3$ (T-phase).

**Table I.** Parameters in Eqs.(9) corresponding to exact solution of the Eqs.(6), and limiting cases[+]

| $\mu$ value | Parameters in Eqs.(9) | | | | |
|---|---|---|---|---|---|
| | amplitude $a_1$ | amplitude $a_2$ | amplitude $b_2$ | width $w$ | shift $x_w$ |
| $\mu \to -1$ | tends to $+\infty$ | tends to $+\infty$ | tends to $-\infty$ | $\sqrt{2}$ | 0, as defined from B.C.[++] |
| $\mu = 0$ | 1 | 1 | 0 | $\sqrt{2}$ | 0, as defined from B.C.[*] |
| $\mu \to 1$ | undefined | undefined | undefined | diverges | undefined[**] |
| $\mu = 3$ | 1 | 0 | 1 | $\sqrt{2}$ | undefined[***] |
| $\mu > 3$ | 1 | 0 | 0 | $\sqrt{2}$ | 0, as defined from B.C. |

[+] The constant amplitude $a_0 \equiv 0$ to satisfy the antisymmetric boundary conditions for $p_1(x)$

[++] The abbreviation "B.C." means boundary conditions

[*] for $\mu = 0$ the equations become decoupled

[**] for $\mu = 1$ the first order phase transition occurs in domain morphology

[***] – for $\mu = 3$ the energy is $x_w$-independent

The variational procedure makes sense only if the trial functions (9) correspond to sufficiently accurate fitting of the numerically calculated domain wall profiles. To verify this, we performed FEM for the case of antiperiodic-periodic boundary conditions (7c) at $l > 100$, and obtained that the functions (9) surprisingly well fit the numerical profiles point-in-point for all $\mu$ values in the range $-1 < \mu < 5$ [compare solid and dashed curves in **Figs. 3**].

From the fitting of FEM results we extracted the variational parameters $a_1$, $a_2$, $b_2$, $w$, and $x_w$, which dependences on $\mu$ are presented in **Fig. 4a-b.** Using the direct variational method, we derived simple analytical expressions for the $\mu$-dependence of the amplitudes $a_1$ and $a_2$:

$$a_1 = \begin{cases} \pm\sqrt{\frac{1}{1+\mu}}, & -1 < \mu < 1, \\ \pm 1, & \mu > 1, \end{cases} \qquad a_2 = \begin{cases} \pm\sqrt{\frac{1}{1+\mu}}, & -1 < \mu < 1, \\ 0, & \mu > 1, \end{cases} \quad (10a)$$

Expressions (10a) are almost exact [see blue and black curves in **Fig. 4a**, where the case corresponding to the sign "+" is shown]. Using the conservation of the first integral at the domain wall and Eqs.(10a), we derived approximate expressions for the amplitude $b_2$:

$$b_2 \approx \begin{cases} \mp\sqrt{\frac{1}{1+\mu}} \mp 1, & -1 < \mu < 1, \\ \pm\frac{x_w}{w}\coth\left(\frac{x_w}{w}\right), & 1 < \mu < 3, \\ 0, & \mu > 3. \end{cases} \quad (10b)$$

The upper signs in the expression (10b) corresponds to red curves in **Fig. 4a**.

The domain wall width, $w$, and shift, $x_w$, have a strong peculiarity in the same region $0.9 < \mu < 1.1$ and can be approximately described by the spline-interpolation functions:

$$w \approx \begin{cases} spline, & -1 < \mu < 1, \\ spline, & 1 < \mu < 3, \\ \sqrt{2}, & \mu > 3, \end{cases} \qquad x_w \approx \begin{cases} 0, & -1 < \mu < 1, \\ spline, & 1 < \mu < 3, \\ 0, & \mu > 3, \end{cases} \quad (10c)$$

Expressions (10c) correspond to brown and green curves in **Fig. 4b**. The variational method determining the analytical dependences of parameters $a_1$, $a_2$, $b_2$, $w$, and $x_w$ on $\mu$ is described in **Appendix B** in Suppl. Mat.[57].

The high accuracy of the fitting results [shown in **Figs. 3-4**] in the entire range $-1 < \mu < 5$, which uses only five $\mu$−dependent parameters for two polarization components, allows us to conclude that the analytical functions (9), which are trial functions in the direct variational method, can be treated as the high-accuracy variational solution of the static EL equations with cubic nonlinearity.

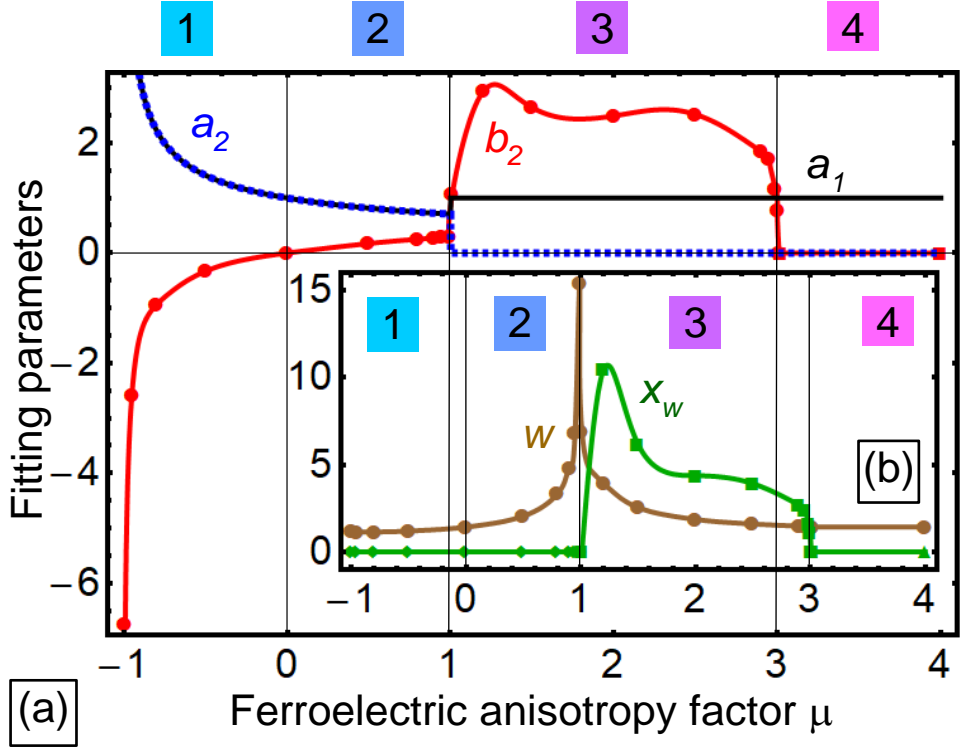

**FIGURE 4. (a)** Dependences of the polarization amplitudes $a_1$, $a_2$ and $b_2$ (solid black, blue dotted and solid red curves respectively). on the control parameter – ferroelectric anisotropy factor y $\mu$. **(b)** Dependences of the shift $x_w$ and width $w$ on the parameter $\mu$ (green and brown curves respectively). Numbers 1 - 4 in the upper row denote regions with different morphology of domain walls, which profiles are shown in **Fig.2**. Regions "1" and "2" correspond to the "dark-dark" and "dark-bright" Ising-Bloch domain walls in O-phase; regions "3" and "4" correspond to the mixed Ising-Bloch and purely Ising domain walls in T-phase. Solid, green, and brown curves are spline-interpolations plotted through the symbols (circles and boxes) calculated by FEM.

## 4. ANALYTICAL POLYDOMAIN SOLUTIONS

### A. Analytical Polydomain Solutions for the "Rotational" Walls

For the case of negative ferroelectric anisotropy factor $\mu \leq -1$ the system of Eqs.(6a) becomes unstable. Here we explore stable polydomain solutions of Eqs.(6) for the case $\mu > -1$ and $l < 100$. After introducing new variables in Eqs.(6), $p = p_1 + p_2$ and $a = p_1 - p_2$, one could get the following equations for them (see Eq. (A.5b) from **Appendix A** in Suppl. Mat.[57]):

$$-p + \frac{1+\mu}{4}p^3 + \frac{3-\mu}{8}p\,a^2 - \frac{\partial^2 p}{\partial x^2} = 0, \tag{11a}$$

$$-a + \frac{1+\mu}{4}a^3 + \frac{3-\mu}{8}a\,p^2 - \frac{\partial^2 a}{\partial x^2} = 0. \tag{11b}$$

It is seen that Eqs.(11a) and (11b) are independent on each other for the specific case $\mu = 3$, when they can be solved separately using elliptic Jacobi functions. The solution for $\mu = 3$ has the form of elliptic sine ("**snoid**") functions:

$$p(x_3) = \sqrt{\frac{2m}{1+m}} sn\left(\frac{x+x_{w1}}{\sqrt{1+m}}\middle| m\right), \quad a(x_3) = \sqrt{\frac{2n}{1+n}} sn\left(\frac{x+x_{w2}}{\sqrt{1+n}}\middle| n\right), \quad (12)$$

where two "modules", $0 \leq m \leq 1$ and $0 \leq m \leq 1$, and two "shifts", $x_{w1}$ and $x_{w2}$, of snoids should be defined from the boundary conditions, as shown below.

Next, using the relations $p_1 = (p+a)/2$ and $p_2 = (p-a)/2$, one obtains from Eqs.(12) the expressions for $p_1$ and $p_2$ in the form of two snoids superposition:

$$p_1(x) = \frac{1}{2}\left[\sqrt{\frac{2m}{1+m}} sn\left(\frac{x+x_{w1}}{\sqrt{1+m}}\middle| m\right) + \sqrt{\frac{2n}{1+n}} sn\left(\frac{x+x_{w2}}{\sqrt{1+n}}\middle| n\right)\right], \quad (13a)$$

$$p_2(x) = \frac{1}{2}\left[\sqrt{\frac{2m}{1+m}} sn\left(\frac{x+x_{w1}}{\sqrt{1+m}}\middle| m\right) - \sqrt{\frac{2n}{1+n}} sn\left(\frac{x+x_{w2}}{\sqrt{1+n}}\middle| n\right)\right]. \quad (13b)$$

Since the solution (13) is dependent on four free parameters, modules $m$ and $n$, and shifts $x_{w1}$ and $x_{w2}$, it pretends to be a general solution, but we cannot say that it is the one, because the existence and uniqueness theorem is not valid for solutions of nonlinear differential equations. The modules $m$ and $n$ define the shape and the period of the polarization profile (13).

A1. Natural and zero boundary conditions for the polarization components. It is shown in **Appendix A** in Suppl. Mat.[57], that the modules, $m$ and $n$, satisfy the same transcendental equations for the two limiting cases of the boundary conditions (7b), $\lambda_i = 0$ and $\lambda_i \to +\infty$, namely:

$$2\sqrt{1+m}\mathbf{K}(m)N_x = l, \qquad 2\sqrt{1+n}\,\mathbf{K}(n)N_y = l. \quad (14a)$$

Here $\mathbf{K}(k)$ is the complete elliptic integral of the first kind, which determines the half-period $2\mathbf{K}(k)$ of the elliptic functions. Also we introduced the numbers $N_x = 0, 1, 2\ldots$ and $N_y = 0, 1, 2, \ldots$ corresponding to the number of "nodes" of $p(x)$ and $a(x)$ functions, which satisfy Eqs.(11a) and (11b), respectively. The situation with these nodes is similar to the eigen solutions of a wave equation, when the boundary conditions are satisfied by an infinite set of solutions with a different number of half-waves for a fixed thickness $l$.

At the same time, we obtained that the shifts $x_{w1}$ and $x_{w2}$ depend on the boundary condition type, namely, and for the natural boundary conditions $\left.\frac{dp_i}{dx}\right|_{x=\pm l/2} = 0$ they have a relatively simple form:

$$x_{w1} = \frac{l}{2}\left(1 \pm \frac{1}{N_x}\right), \quad x_{w2} = \frac{l}{2}\left(1 \pm \frac{1}{N_y}\right), \quad (\lambda_i \to +\infty). \quad (14b)$$

For zero polarization conditions $p_i|_{x=\pm l/2} = 0$ the shifts are given by more complex expressions:

$$x_{w1} = \frac{l}{2}\left(1 \pm \frac{1-(-1)^q}{N_x}\right), \quad x_{w2} = \frac{l}{2}\left(1 \pm \frac{1-(-1)^s}{N_y}\right), \quad (\lambda_i = 0), \quad (14c)$$

where $q$ and $s$ are independent integers. From Eqs.(14) the par of integers $\{N_x, N_y\}$ characterizes the domain structure of the solution (13) at a given $l$. However, the characterization is not unique due to the two possible signs "±" and different powers $q$ and $s$ in Eqs.(14b)-(14c).

The case $N_x = 0$ (or $N_y = 0$, or both) requires a separate consideration, since it corresponds to the limiting cases $m \to 1$ (or/and $n \to 1$), respectively. The shift $x_{w1,2}$ diverge as $-\log(\sqrt{1-k})$ at $k \to 1$ ($k = m, n$) and the solutions (12)-(13) formally becomes undefined. More rigorous consideration shows that the case corresponds to "nodeless" solutions, which are trivial and independent on the film thickness

$$p(x_3) \xrightarrow{N_x \to 0} \begin{cases} 0, & \lambda_i = 0, \\ 1, & \lambda_i \to +\infty. \end{cases} \qquad a(x_3) \xrightarrow{N_y \to 0} \begin{cases} 0, & \lambda_i = 0, \\ 1, & \lambda_i \to +\infty. \end{cases} \qquad (14d)$$

Using the "decoupled" form (12) of the solution (13) in a general case, the "decoupled" free energy, $G[p, a]$, is equal to:

$$G[p, a] = \frac{1}{l} \int_{-\frac{l}{2}}^{\frac{l}{2}} dx \left[ -\frac{p^2 + a^2}{4} + \frac{p^4 + a^4}{8} + \frac{1}{4}\left(\left(\frac{dp}{dx}\right)^2 + \left(\frac{da}{dx}\right)^2\right) \right] \equiv G[m, n], \quad (15a)$$

$$G[m, n] = G_c[m] + G_c[n], \qquad G_c[k] = -\frac{1}{6(1+k)^2}\left(2 + k - 2(1+k)\frac{\mathbf{E}(k)}{\mathbf{K}(k)}\right), \quad (15b)$$

where $\mathbf{E}(k)$ is the complete elliptic integral of the second type. Since the energy is independent on $x_{w1,2}$, it is the same for both cases $\lambda_i = 0$ and $\lambda_i \to +\infty$. The case $m \to 1$ (or $n \to 1$) is exceptional, and corresponding contributions to Eq.(15a) should be rewritten as follows:

$$G_c[m] \xrightarrow{N_x \to 0} \begin{cases} 0, & N_x \to 0, \lambda_i = 0; \\ -\frac{1}{8}, & N_x \to 0, \lambda_i \to +\infty. \end{cases} \qquad (15c)$$

The normalized energy $G[m, n]$ of the polydomain states as a function of $l$ was calculated from Eqs.(15) for different numbers of domain walls inside the film, $\{N_x, N_y\}$. Results are shown in **Fig. 5a** for $\lambda_i \to +\infty$, and in **Fig.6b** for $\lambda_i = 0$. The energy is normalized on a "bulk" value $G_b = 1/4$. Typical distribution of polarization components, $p_1$ (red curves) and $p_2$ (blue curves) calculated with Eqs.(13) for different $N_x$ and $N_y$, are shown in **Figs. 5b-g** for $\lambda_i \to +\infty$, and in **Fig.6b-g** for $\lambda_i = 0$.

The metastable and stable polydomain states have negative energy, which monotonically decreases with $l$ increase for a fixed $\{N_x, N_y\}$ in both cases $\lambda_i \to +\infty$ and $\lambda_i = 0$ (see **Fig.5a** and **6a**). The single-domain state with $N_x = N_y = 0$ is absolutely stable for $\lambda_i \to +\infty$ (corresponding energy relief corresponds to the potential well) and unstable for $\lambda_i = 0$ (corresponding energy relief corresponds to the saddle point) (compare horizontal dashed lines in **Fig.5a** and **6a**). The critical thickness of the film, $l_{cr}$, below which the ferroelectric phase disappears, is individual for the concrete polydomain state $\{N_x, N_y\}$, and, as a rule, it increases with sum $N_x + N_y$ increase (see vertical dotted lines in **Fig.5a** and **6a**). Only the single-domain state {0,0} has no critical thickness at $\lambda_i \to +\infty$. The states {0,1}, {1,1}, {1,2}, {1,3} ... and {1,$N$} have the same minimal $l_{cr} \approx \pi$. The states {0,2}, {2,2}, {2,3}, ... and {2,$N$} have the same $l_{cr} \approx 2\pi$. The states {0,3}, {3,3}, {3,4}, ... and {3,$N$} have the same $l_{cr} \approx 3\pi$. In general, all states with the same $N = min\{N_x, N_y\}$ have the same $l_{cr}[N]$, which increases with $N$ increase. The critical thickness can be derived from Eqs.(14a) in the limit $m \to 0$ (or $n \to 0$), namely:

$$l_{cr}[N] \approx \pi N \quad (N > 0). \quad (16)$$

For a fixed $l$ and $\lambda_i \to +\infty$ the energy of the polydomain states increases with the sum $N_x + N_y$ increase (see **Figs. 5a**), and the lowest polydomain state is $\{0,1\} = \{1,0\}$. The energy of {0,2} state is slightly lower that the energy of {1,1} state, but this state has twice bigger $l_{cr}$. For $l > l_{cr}$ the energies of {1,1} and {0,2} states become very close and approach the energy of the single-domain state in the limit $l \gg l_{cr}$. The same trend is evident for all other states $\{N_x, N_y\}$ and $\{0, N_x + N_y\}$. It is important that the energy of all polydomain states tends to the single-domain state energy $G = -G_b$ in the limit $l \to +\infty$ for the case $\lambda_i \to +\infty$.

It is seen from **Fig. 5b-g** that the profile and amplitude of the polydomain solution (13) is determined by the film thickness $l$ and by the "nodes" pair $\{N_x, N_y\}$. Simple Ising-type domains with quasi-sinusoidal profile for $l$ slightly bigger than $l_{cr}$ (see **Figs. 5b**), or with strongly anharmonic "snoidal" profile for $l \gg l_{cr}$ (see **Figs. 5e**), correspond to the same numbers $N_x = N_y$. Rather complex phase-shifted asymmetric Bloch-Ising type profiles, which are quasi-harmonic for $l \approx l_{cr}$ (see **Figs. 5c**), and strongly anharmonic for $l \gg l_{cr}$ (see **Figs. 5f**), correspond to the close numbers $N_x = N_y - 1$. Simple in-phase [i.e. $p_1(x) = p_2(x) + 1$] and symmetric [i.e. $p_{1,2}(x) = p_{1,2}(-x)$] Bloch-Ising type profiles, which are quasi-harmonic for $l \approx l_{cr}$ (see **Figs. 5d**), and strongly anharmonic for $l \gg l_{cr}$ (see **Figs. 5g**), correspond the pairs $\{0, N_y\}$ or $\{N_x, 0\}$.

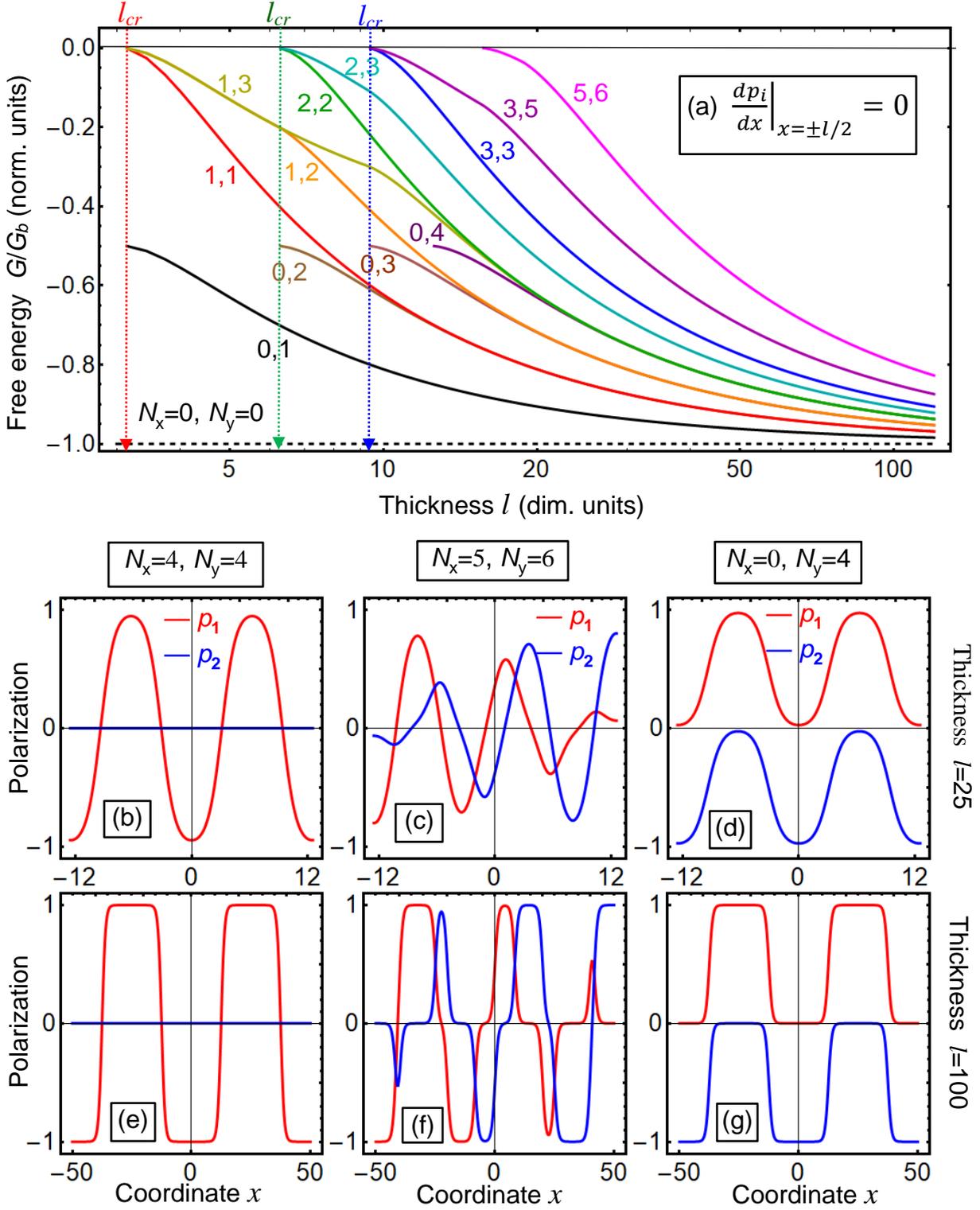

**FIGURE 5.** (a) Energy of the metastable polydomain states as a function of dimensionless film thickness $l$ for different numbers of domain walls inside the film, $\{N_x, N_y\}$, denoted near the curves. (b-g) Distribution of polarization components, $p_1$ (red curves) and $p_2$ (blue curves), in the film of thickness $l = 25$ (b, c, d) and $l = 100$ (e, f, g) calculated using Eqs.(13) for different numbers $N_x=4$ and $N_y=4$ (b, e); $N_x=5$ and $N_y=6$ (c, f); $N_x=0$ and $N_y=4$ (d, g). The absolutely stable single-domain state with $N_x = N_y = 0$ is shown for comparison

by a dashed line. Natural boundary conditions are used for polarization components: $\frac{dp_i}{dx}\big|_{x=\pm l/2} = 0$, i.e. $\lambda_i \to +\infty$ for all plots **(a-g)**. Ferroelectric anisotropy factor $\mu = 3$, and energy scale factor $G_b = 1/4$.

For a fixed $l$ and $\lambda_i = 0$ the energy of the polydomain states increases with $N_x$ or/and $N_y$ increase (see **Figs. 6a**), and the lowest polydomain states are $\{1,1\}$, $\{1,2\}$ and $\{1,3\}$, respectively. The energy of $\{0,1\}$ state is almost the same as the energy of $\{1,3\}$ state only for $l < 3\pi$; at $l \gg 3\pi$ it tends to $-G_b/2$, while the energy of the $\{1, N_y\}$ states tends to $-G_b$ in the limit $l \gg l_{cr}$. The same trend is evident for all other states $\{N_x, N_y\}$ ($N_{x,y} \geq 1$), which energy eventually tends to $-G_b$ in the limit $l \to +\infty$. The energy of $\{0, N_y\}$ eventually tends to $-G_b/2$ in the limit $l \to +\infty$. It is important that in the limit $l \to +\infty$ the energies $G$ of all polydomain states "split" into two levels – the ground domain state "0" with $G_0 = -G_b$ and the excited state "1" with $G_1 = -\frac{G_b}{2}$, which are separated by the "gap" of width $\Delta G = \frac{G_b}{2}$. These two levels, each of which splits on the infinite set of sub-levels, which are characterized by a multiple close-energy polydomain morphologies with number $\{0, N_y\}$ and $\{N_x, N_y\}$ ($N_{x,y} \geq 1$), respectively.

The two-level energy structure of the polydomain states for $\lambda_i = 0$ principally differs from the single level existing in the case $\lambda_i \to +\infty$ (compare **Fig.5a** and **6a**). Since zero polarization at the film surfaces can be realized experimentally by creation of the ultra-thin non-ferroelectric passive layers at the surfaces [59], this suggests possible strategies to switch the polarization state between the sub-levels "0" and "1". Imagine that we have excited (e.g. by electric field) the film polarization to the one of the polydomain states "1". When the system is released, it tries to thermalize its energy excess, and, if the dissipation is very small in a film without imperfections [i.e. $\rho \gg \tau^2$ in Eqs.(6)], it can oscillate with some period between the excited states "1" and ground state "0".

It is seen from **Fig. 6b-g** that the profile and amplitude of the polydomain solution (13) is determined by the film thickness $l$ and by the "nodes" pair $\{N_x, N_y\}$, but the details of domain pattern slightly differs from the ones shown in **Fig. 5b-g**. Simple Ising-type domains with quasi-sinusoidal profile for $l$ slightly bigger than $l_{cr}$ (see **Figs. 6b**), or with strongly anharmonic "snoidal" profile for $l \gg l_{cr}$ (see **Figs. 6e**), correspond to the same numbers $N_x = N_y$. Rather complex phase-shifted asymmetric Bloch-Ising type profiles, which are quasi-harmonic for $l \approx l_{cr}$ (see **Figs. 6c**), or strongly anharmonic for $l \gg l_{cr}$ (see **Figs. 6f**), correspond the close numbers $N_x = N_y - 1$. Simple anti-phase [i.e. $p_1(x) = -p_2(x)$] and antisymmetric [$p_{1,2}(x) = -p_{1,2}(-x)$] Bloch-Ising type profiles, which are quasi-harmonic for $l \approx l_{cr}$ (see **Figs. 6d**), and strongly anharmonic for $l \gg l_{cr}$ (see **Figs. 6g**), correspond the pairs $\{0, N_y\}$ or $\{N_x, 0\}$.

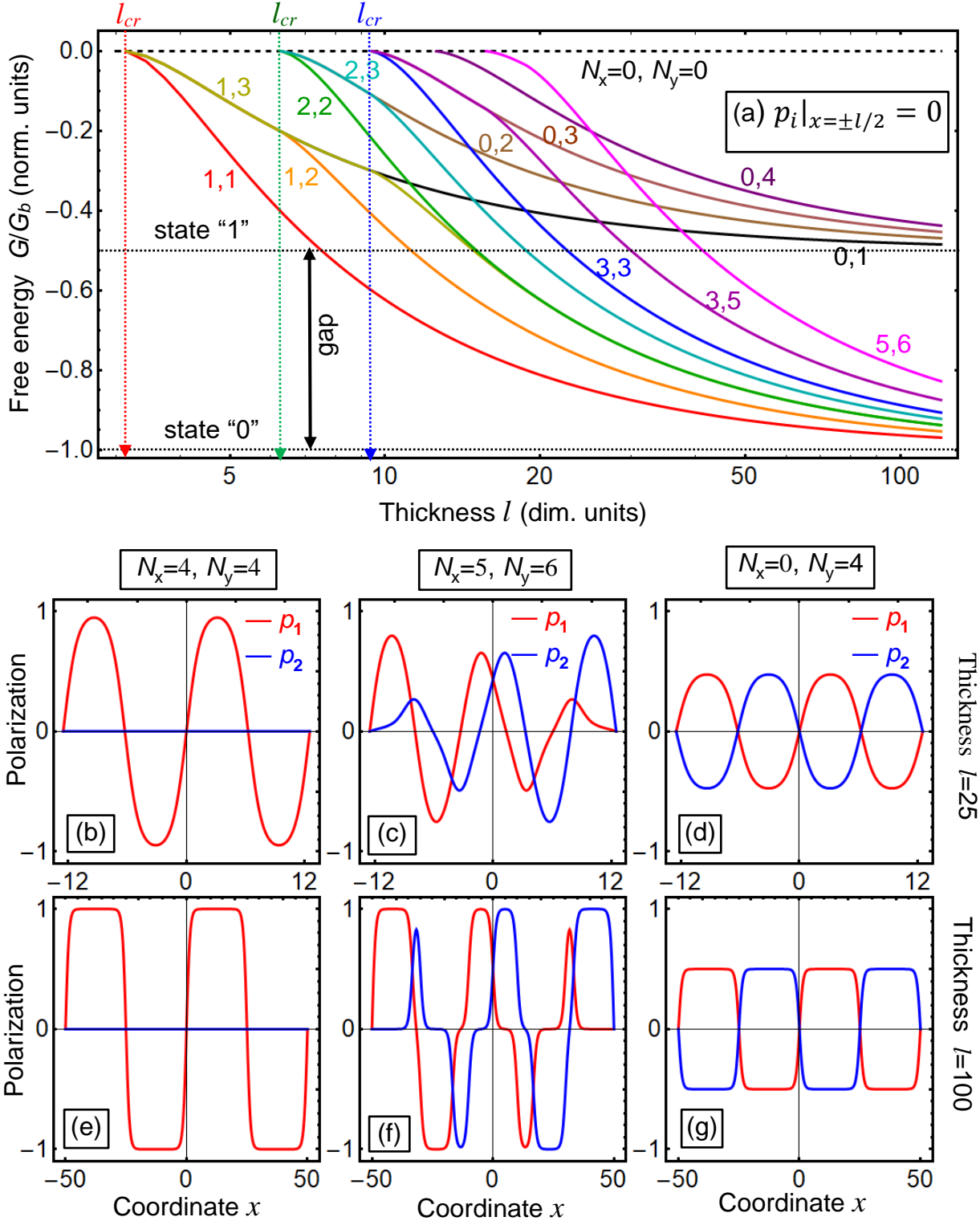

**FIGURE 6.** (a) Energy of the polydomain states as a function of dimensionless film thickness $l$ for different numbers of domain walls inside the film, $\{N_x, N_y\}$, denoted near the curves. (b-g) Distribution of polarization components, $p_1$ (red curves) and $p_2$ (blue curves), in the film of thickness $l = 25$ (b, c, d) and $l = 100$ (e, f, g) calculated using Eqs.(13) for different numbers $N_x=4$ and $N_y=4$ (b, e); $N_x=5$ and $N_y=6$ (c, f); $N_x=0$ and $N_y=4$ (d, g). The unstable single-domain state with $N_x = N_y = 0$ is shown for comparison by a dashed line.

Ferroelectric anisotropy factor $\mu = 3$, and $G_b = 1/4$. Zero boundary conditions are used for polarization components, $p_i|_{x=\pm l/2} = 0$, i.e. $\lambda_i = 0$ for all plots **(a-g)**.

To resume the analysis of the limiting cases $\lambda_i \to \infty$ and $\lambda_i = 0$, analytical solutions (13), which contain 4 free parameters ($m, n, x_{w1}$ and $x_{w2}$), are suitable candidates for a general equilibrium solution of nonlinear differential EL equations (6) for $\mu = 3$, since the number of free parameters is enough to satisfy arbitrary boundary conditions (7) at the film surfaces. The solutions (13) are degenerated for a fixed boundary conditions, because they contain different number of domain walls proportional to $N_x$ for $p_1 + p_2$, and to $N_y$ for $p_1 - p_2$. The analysis of the free energy (15) dependence on the number of domains for a fixed film thickness $l$ allows to select the single- or polydomain solution corresponding to the minimal energy [see **Fig. 5a** and **6a**]. We obtained that the single-domain state corresponds to the minimal energy for the case of zero polarization derivative at the film surfaces [namely for $\lambda_i \to \infty$ in Eq.(7b)], while the Ising-Bloch polydomain states with the total number of rotational domain walls $N_x + N_y \geq 1$ minimize the system energy for zero polarization at the film surfaces [namely for $\lambda_i = 0$ in Eq.(7b)]. In the case $\lambda_i = 0$ the energy of polydomain states splits into two levels "0" and "1", and each of levels is the great number of the close-energy sub-levels (infinite in the limit $l \to \infty$), which domain structure is characterized by the pair of nodes $\{0, N_y\}$ for the level "1" and $\{N_x, N_y\}$ for the level "0", where $N_{x,y} \geq 1$.

A2. Periodic-antiperiodic boundary conditions for the polarization components. If the periodic-antiperiodic boundary conditions (7c) are valid for two components of polarization, one can put $m = n$ and $x_{w1} = -x_{w2} = x_w$ in the solution (13), while it is not the only possibility in the case. Anyway, the two parameters, $m$ and $x_w$, remained undefined from the Eqs.(13) in the case of Eqs.(7c). The free energy $G$ is minimal for $m \to 1$ and $n \to 1$ corresponding to a solitary domain wall. In the simultaneous limit $m=n \to 1$ the solution (13) transforms into a solitary wall solution (9), namely $p_1(x_3) = \frac{1}{2}\left[\tanh\left(\frac{x+x_w}{\sqrt{2}}\right) + \tanh\left(\frac{x-x_w}{\sqrt{2}}\right)\right]$ and $p_2(x_3) = \frac{1}{2}\left[\tanh\left(\frac{x+x_w}{\sqrt{2}}\right) - \tanh\left(\frac{x-x_w}{\sqrt{2}}\right)\right]$. These expressions become identical with Ivanchik solution [39] after elementary calculations (see **Appendix A** in Suppl. Mat.[57] for details).

The solution (13) can be considered as a trial function for $-1 < \mu < 3$. For instance, in the case of the boundary conditions (7c) the trial functions for polydomain solutions are

$$p_1(x) = a_1 + \frac{b_1}{2}\sqrt{\frac{2m}{1+m}}\left[sn\left(\frac{x+x_w}{\sqrt{1+m}}\bigg| m\right) + sn\left(\frac{x-x_w}{\sqrt{1+m}}\bigg| m\right)\right], \quad (17a)$$

$$p_2(x) = a_2 + \frac{b_2}{2}\frac{w}{x_w}\sqrt{\frac{2m}{1+m}}\left[sn\left(\frac{x+x_w}{\sqrt{1+m}}\bigg| m\right) - sn\left(\frac{x-x_w}{\sqrt{1+m}}\bigg| m\right)\right], \quad (17b)$$

where "free" parameters, constant offsets $a_i$ and amplitudes $b_i$, are introduced in the same way as in Eqs.(9). They should be determined by using the direct variational method similarly to the case of a solitary domain wall considered in the previous **Section 3**.

## B. Analytical Solutions for the Ising Polydomains

For the case of high positive ferroelectric anisotropy $\mu > 3$ the stable polydomain solution of Eqs.(6a) becomes of Ising type and $\mu$-independent, since $p_2$-component is absent. Corresponding polarization profile has the snoidal form:

$$p_1(x_3) = \sqrt{\frac{2m}{1+m}} \, \text{sn}\left(\frac{x+x_w}{\sqrt{1+m}}\bigg| m\right), \qquad p_2(x_3) \equiv 0. \qquad (18a)$$

The snoid modulus $m$ and shift $x_w$ should be determined from the boundary conditions (7). For the two limiting cases, $\lambda_i = 0$ and $\lambda_i \to +\infty$, the modulus $m$ satisfies the condition $2N\sqrt{1+m}\,\mathbf{K}(m) = l$, where the integer number $N$ regulates the number of solution "nodes" [compare with Eqs.(14a)]. The shift $x_w = \frac{l}{2}\left(1 \pm \frac{1}{N}\right)$ for $\lambda_i \to +\infty$, or $x_w = \frac{l}{2}\left(1 \pm \frac{1-(-1)^s}{N}\right)$ for $\lambda_i = 0$ [compare with Eqs.(14b)-(14c)]. Using the solution (18a), the free energy was derived as:

$$G[m] = \frac{1}{l}\int_{-\frac{l}{2}}^{\frac{l}{2}} dx \left(-\frac{p^2}{2} + \frac{p^4}{4} + \frac{1}{2}\left(\frac{dp}{dx}\right)^2\right) \equiv -\frac{1}{3(1+m)^2}\left(2 + m - 2(1+m)\frac{\mathbf{E}(m)}{\mathbf{K}(m)}\right). \qquad (18b)$$

The first integral is $I(m) = -\frac{m}{(1+m)^2}$. As it can be seen, Eqs.(18) are the particular case of Eqs.(13)-(15) for the case $N_x = N_y = N$. Thus, the solutions (18a) and their energy (18b) are presented among other curves with $N_x = N_y$, which are shown in **Figs.5-6.** To resume, the analytical polydomain solution (18a), which contain 2 free parameters ($m$ and $x_{w2}$), is suitable candidates for a general stable solution of EL equations (6) for $\mu > 3$, since the number of free parameters is enough to satisfy the boundary conditions (7) at the film surfaces. The solution (18) is degenerated for a fixed boundary conditions, because it contains different number of domains ($N$ or $N+1$). The analysis of the free energy (18b) dependence on the number of domains for a fixed film thickness $l$ allows to select the single- or polydomain solution corresponding to the minimal energy [see curves {1,1}, {2,2} and {3,3} in **Figs. 5a** and **6a**]. The single-domain state corresponds to the minimal energy for the case of polarization zero derivative at the surface [namely for $\lambda_i \to \infty$ in Eq.(7b)], while the Ising polydomains minimize the system energy for e.g. zero polarization at the surface [namely for $\lambda_i \to 0$ in Eq.(7b)].

# 5. POSSIBLE APPLICATIONS OF ANALYTICAL RESULTS FOR BAYESIAN ANALYSIS AND INFORMATION PROCESSING

## A. Bayesian Analysis of Domain Walls Profiles

Let us consider a model situation when we know the ferroelectric film thickness $L$, the temperature $T$ is fixed, and the most of ferroelectric material parameters in Eqs.(1) are determined from independent experiments, or tabulated. For instance, the parameters $a_i$, $a_{ii}$ and $a_{ij}$ can be determined from the measurements of dielectric permittivity and spontaneous polarization temperature dependences in a bulk homogeneous ferroelectric [4, 5]. Much more complex is to determine the gradient coefficients, $g_{ijkl}$, and the error originated from $g_{ijkl}$ estimation using the intrinsic width of differently-oriented uncharged domain walls [4, 5] is typically high (~ 50 – 100%) due to wall pinning and "trembling" near lattice defects and other imperfections. Moreover, the direct determination of extrapolation lengths $\lambda_i$, introduced in Eq.(7b), is a true challenge, because the surface energy coefficients $a_i^S$ in Eq.(1b) depend on the surface/interface chemistry and surface defects in a prior unknown way, and only indirect estimates have been done for several particular cases [58]. The only exception may be the deposition of artificial dead layers [59] at the film surfaces, which yield $p_i(\pm l/2) = 0$ and so $\lambda_i = 0$. Hence, the dimensionless control parameter $\mu$ can be regarded known for a fixed temperature, while the dimensionless extrapolation and correlation lengths, $\lambda_i \cong \frac{g_{44} R_c}{a_i^S}$ and $R_c \cong \sqrt{g_{44}/|2a_1|}$, are *prior* unknown.

In the previous section we calculated the energy levels, $G[m,n]$, given by Eqs.(15). The re-definition $G[m,n] \equiv G(N_x, N_y)$ is possible, since the relation between the pair of modules $\{m,n\}$ and integers $\{N_x, N_y\}$ determining the domain numbers is established by e.g. Eq.(14a) for $\mu = 3$ in particular cases $\lambda_i \to +\infty$ and $\lambda_i \to 0$. Since these results can be extended for the case of arbitrary $\mu > -1$ by application of direct variational method, as described above, we can assume that the energy levels describing different polarization states $p_i(x, N_x, N_y, R_c, \lambda)$, the expected (i.e. averaged over all realizations) polarization state $\langle p_i(x, R_c, \lambda) \rangle$ and energy $\langle G(R_c, \lambda) \rangle$ can be calculated for the *posterior* known sets of parameters $\lambda = \{\lambda_i\}$ and $R_c$. Namely:

$$\langle p_i(x, R_c, \lambda) \rangle = \sum_{N_{x,y}=0}^{N_{x,y}=N_m} p_i(x, N_x, N_y, R_c, \lambda) \cdot W[N_x, N_y | R_c, \lambda], \qquad (19a)$$

$$\langle G(R_c, \lambda) \rangle = \sum_{N_{x,y}=0}^{N_{x,y}=N_m} G(N_x, N_y, R_c, \lambda) \cdot W[N_x, N_y | R_c, \lambda], \qquad (19b)$$

where $i = 1,2$, and we introduced the designation $\lambda = \{\lambda_i\}$ for the sake of brevity. The maximal number of the energy states $N_m$ is defined from the condition $N_m = Trunc\left[\frac{l}{\pi}\right]$, obtained from the Eq.(16) for $l_{cr}(N_m)$. For example, $N_m = 6$ for $l = 20$ (see **Fig. 5a** and **6a**).

The conditional prior probability $W[N_x, N_y | g, \lambda]$ of the state $p_i(x, N_x, N_y, R_c, \lambda)$ realization under the validity of "event" – realization of the parameters $\{R_c, \lambda\}$ in the film, is expressed via the energy levels $G(N_x, N_y, R_c, \lambda)$ and canonic statistic sum $Z(g, \lambda)$ in a conventional way:

$$W[N_x, N_y | R_c, \lambda] = \frac{k_{N_{x,y}}}{Z(R_c, \lambda)} exp\left[-\frac{G(N_x, N_y, R_c, \lambda)}{k_B T}\right], \quad (19c)$$

$$Z(R_c, \lambda) = \sum_{N_{x,y}=0}^{N_{x,y}=N_m} G(N_x, N_y, R_c, \lambda). \quad (19d)$$

Here the statistical weight $k_{N_{x,y}}$ the of the state $p_i(x, N_x, N_y, R_c, \lambda)$ is equal to 2 (when degeneration $p_1 \to -p_1$ exists) or 4 (when degeneration $p_1 \to \pm p_2$ exists) depending on $\mu$-value.

Using the Bayesian approach described in Refs.[54, 55], we can determine the posterior conditional probability $W[R_c, \lambda | N_x, N_y]$ of the $\{R_c, \lambda\}$ parameters:

$$W[R_c, \lambda | N_x, N_y] = \frac{W[N_x, N_y | R_c, \lambda]}{W[N_x, N_y]} W[R_c, \lambda], \quad (20a)$$

where the "***marginal***" or "***unconditional***" probability $W[N_x, N_y]$ of the state $p_i(x, N_x, N_y)$ realization is the sum and/or integral over all prior probabilities $W[R_c, \lambda]$ of the $\{R_c, \lambda\}$-events:

$$W[N_x, N_y] = \int_{R_{min}}^{R_{max}} dR_c \int_0^\infty d\lambda\, W[R_c, \lambda] W[N_x, N_y | R_c, \lambda] \quad (20b)$$

In practice one can determine the probability $W[N_x, N_y]$ by the fitting of analytical profiles (17) to the domain wall profiles, which are measured experimentally. Thus, expressions (20) illustrate that the energies (15), corresponding to the analytical solutions (13) or (17), and statistical probabilities (19) can be used for a Bayesian analysis of domain wall profiles reconstructed from atomic displacements measured by HR STEM in thin ferroelectric films (see Refs.[54, 55])

As a toy model, let us assume that $R_c$ is determined reliably from the intrinsic width of domain walls, and consider only 2 possible and equiprobable events, $\lambda = +\infty$ and $\lambda = 0$. Under these conditions, Eqs.(20) acquire much simpler form:

$$W[\lambda | N_x, N_y] \cong \frac{W[N_x, N_y | \lambda]}{W[N_x, N_y | 0] + W[N_x, N_y | \infty]}, \quad (21a)$$

where $\lambda = +\infty$ or $\lambda = 0$.

Alternatively, when all $\lambda$ values can be regarded quasi-equiprobable in the region $\{0, \lambda_{max}\}$, we can put $W[R_c, \lambda] \approx \frac{1}{\lambda_{max}}$ and simplify Eqs.(20) as follows:

$$W[\lambda | N_x, N_y] \cong \frac{W[N_x, N_y | \lambda]}{\int_0^{\lambda_{max}} d\lambda\, W[N_x, N_y | \lambda]}, \quad (21b)$$

Expressions (21) demonstrates that the Bayes' formula allows us to estimate the posterior probability of different $\lambda$ values realization, since the posterior probability is proportional to the prior one. The tree-like diagram illustrating the application of Bayes' formula for Eqs.(21) is shown in **Fig. 7**. The diagram demonstrates possible paths between the prior guess for extrapolation length ($\lambda=0$ or $\lambda=\infty$) and posterior measurement of the domain state corresponding to the sub-levels "0" with domain numbers $\boldsymbol{L_0} = \{N_x, N_y\}$, or sub-levels "1" with domain numbers $\boldsymbol{L_1} = \{0, N_y\} \cup \{N_x, 0\}$, where $N_{x,y} \geq 1$ (see **Figs. 5a** and **6a** for details).

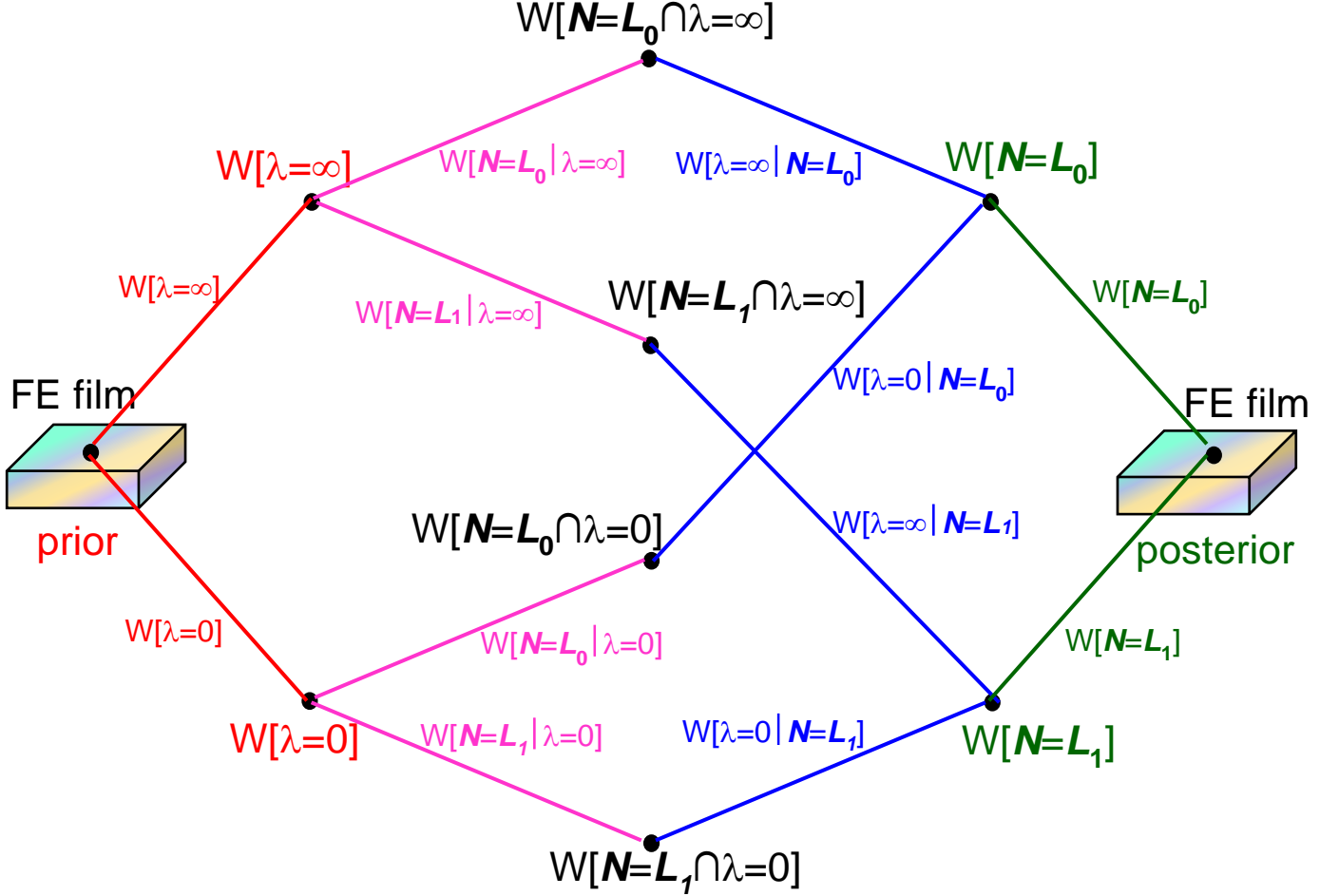

**FIGURE 7.** The tree-like diagram illustrating the Bayes' formula for Eqs.(21), where we introduce the domain number $\boldsymbol{N} = \{N_x, N_y\}$, and the designations for domain structure of the energy levels "0" and "1", $\boldsymbol{L_0} = \{N_x, N_y\}$ and $\boldsymbol{L_1} = \{0, N_y\} \cup \{N_x, 0\}$, respectively ($N_{x,y} \geq 1$). Other designations: W[A] is a probability of the event A, W[B|A] is the probability of event B under the condition A, and W[A∩B] is a mutual probability of event AB. All paths between the prior guess for extrapolation length ($\lambda=0$ or $\lambda=\infty$) and posterior measurement of domain numbers ($\boldsymbol{N} = \boldsymbol{L_0}$ or $\boldsymbol{N} = \boldsymbol{L_1}$) illustrate the symmetry of Bayes' formula $W[B|A]\,W[A] = W[A \cap B] = W[A|B]\,W[B]$.

### B. Information Processing

Let us consider a rather thick ferroelectric film ($l \gg 100$) with zero polarization at its surfaces ($p_i|_{x=\pm l/2} = 0$), which versatile domain morphology is determined by the maximal number of

domain walls $N_m \gg 1$, where $N_m$ is defined from the condition $N_m = Trunc\left[\frac{l}{\pi}\right]$. Corresponding energy structure is similar to the right part of the two-level system shown in **Fig. 6a**, at that the level "1" is divided on $2N_m$ sub-levels with a node structure $\{0, N_y\} \cup \{N_x, 0\}$; and the level "0" is divided on $N_m^2$ sub-levels with a node structure $\{N_x, N_y\}$, where $N_{x,y} \geq 1$. The energies of the sub-levels "1" are very close to $G_1 = -1/2$, and the energies of all sub-levels "0" are very close to $G_0 = -1$ (in $G_b$ units). "Very close" means that the energy difference between the sub-levels $\Delta G_{x,y}$ is smaller than thermal fluctuations energy, $k_B T$, and decreases with $l$ increase. So that each of the complex levels "0" and "1" presents itself a great number of polarization states, which are almost indistinguishable at $k_B T \geq \Delta G_{x,y}$. However, since each sub-level is characterized by a different number (and often configuration) of domain walls, the two-level system can be considered as a very big multi-bit that potentially can imitate a quantum bit (**q-bit**) the better, the larger is the number $N_m$.

Let us imagine that we initially excite the film polarization to the one of the $2N_m$ polydomain states "1". When the system is released, it tries to thermalize its energy excess, and, if the dissipation is very small in a film without imperfections [i.e. $\rho \gg \tau^2$ in Eqs.(6)], it can oscillate with some period between the excited states "1" and the ground states "0". Since each level has the great number of sub-levels, the oscillations between the levels can be used for e.g. racetrack memory and q-bit operation imitation, which utilizes the parallelism of the polarization states evolution.

To elaborate the idea, we preliminary study the oscillatory dynamic of domain structure under the absence and in the presence of losses, for initial seeding in the form of a random polarization distribution and for the boundary condition $p_i|_{x=\pm l/2} = 0$. We obtained that in some cases the polarization vector oscillates (with or without damping in dependence on the loss factor $\tau$) between several metastable domain states, without smearing to all available states. This result can be considered as the direct demonstration of the Fermi-Pasta-Ulam-Tsingou (FPUT) paradox [60] (other names - FPU problem or recurrence [61]), which states that the energy in a weakly coupled non-linear system avoids thermalization and travels between the available modes of the system without diffusing to all available modes.

## 6. SUMMARY

We considered the dynamics of a 180-degree uncharged rotational domain wall in a multiaxial ferroelectric film within the framework of analytical LGD approach. FEM was used to solve numerically the system of the coupled nonlinear EL differential equations of the second order for two components of polarization. It appeared, that the static wall structure (e.g. Ising, Ising-Bloch, or mixed type) and corresponding (meta)stable phase of the film are dependent on the single control parameter – dimensionless factor of ferroelectric anisotropy $\mu$, that can vary in a continuous range $-1 < \mu$. Using spline-interpolations, we fitted the static profile of a solitary domain wall, calculated by FEM, with multi-parametric hyperbolic kink-like functions for the two-component polarization, and extracted the five $\mu$-dependent parameters from the fitting to FEM curves. The surprisingly high accuracy of the fitting results in the entire range $-1 < \mu < 5$, allows concluding that the analytical functions, which are trial functions in the direct variational method, can be treated as the high-accuracy variational solution of the static EL equations with cubic nonlinearity.

Next, using LGD approach, we derived and analyzed the two-component and one-component analytical solutions of the static EL-equations for a polydomain 180-degree domain structure in a multiaxial ferroelectric film. The analytical solutions in the form of elliptic Jacobi functions, which contain 2 and 4 free parameters, respectively, are suitable candidates for a general stable solution of EL equations, since the number of free parameters is enough to satisfy the wide class of boundary conditions at the film surfaces The solutions are degenerate for definite boundary conditions, because they can contain different number of domains for $p_1$, and $p_2$ components. However, the analysis of the free energy dependence on the number of domains for a fixed film thickness allowed us to select those single- or polydomain analytical solution, which corresponds to the minimal energy. In particular, we obtained that the single-domain state corresponds to the minimal energy for the case of the polarization zero derivative at the film surfaces, while the solution with Ising-Bloch domain walls minimize the system energy for zero polarization at the film surfaces. The analytical solutions can become a useful tool for Bayesian analysis of domain wall profiles reconstructed from atomic displacements measured by HR STEM in ferroelectric films [54, 55].

For thick films with zero polarization at the surfaces the energy of the polydomain states split into two levels "0" and "1". Each of the levels "0" or "1" contains a large number of the close-energy sub-levels, which domain morphology is characterized by different structure of nodes for the two-component polarization $\boldsymbol{p} = \{p_1, p_2\}$, namely $\{0, N_y\}$ for the level "1" and $\{N_x, N_y\}$ for the level "0", where $1 \leq N_{x,y} \leq N_m$ and $N_m \gg 1$. Since zero polarization at the surface can be realized experimentally relatively easily by creation of sub-surface non-ferroelectric passive layers, one can switch the polarization state between the levels "0" and "1". Under certain favorable conditions, the two-level system can oscillate with some period between the excited states "1" and ground state "0", and the oscillations can be used for e.g. racetrack memory and q-bit operation imitation.

**Acknowledgements.** Authors gratefully acknowledge Bobby G. Sumpter (ORNL) for useful suggestions and improvements of the manuscript. A.N.M. acknowledges the Target Program of Basic Research of the National Academy of Sciences of Ukraine "Prospective basic research and innovative development of nanomaterials and nanotechnologies for 2020 - 2024", Project № 1/20-Н (state registration number: 0120U102306). This effort (SVK) is based upon work supported by the U.S. Department of Energy (DOE), Office of Science, Basic Energy Sciences (BES), Materials Sciences and Engineering Division and was performed at the Oak Ridge National Laboratory's Center for Nanophase Materials Sciences (CNMS), a U.S. Department of Energy, Office of Science User Facility.

**Authors' contribution.** A.N.M. and S.V.K. generated the research idea, stated the problem and wrote the manuscript. A.N.M and E.A.E. performed analytical calculations. E.A.E. and Y.M.F. wrote the codes and performed numerical calculations.

# SUPPLEMENTARY MATERIALS

## APPENDIX A. Static Solutions

### AI. Analytical solutions in the form of one-component solitary domain walls

For the case $a_{12} = -2a_{11}$ the solitary solution of Eqs.(3) becomes one-component and unstable, namely, $P_1(x_3) = P_2(x_3) = A \sin\left[\frac{x_3-x_0}{R_c}\right]$. For the case $a_{12} = 0$ the solution becomes "decoupled" and have the form of two independent tanh-profiles,

$$P_1(x_3) = P_S \tanh\left[\frac{x_3-x_L}{R_c}\right], \quad P_2(x_3) = P_S \tanh\left[\frac{x_3-x_R}{R_c}\right]. \tag{A.1}$$

For the specific case $a_{12} = 6a_{11}$, $g_{44} > 0$ and $a_1 < 0$ the partial solution of Eqs.(9) was derived by Ivanchik. It presents a an uncharged Ising-Bloch type domain wall (two "rotational" 180-degree c-domains separated by an a-domain).

### AII. Analytical solutions in the form of the two-component Ising-Bloch walls

Adding and subtracting Eqs.(3a) and (3b) from one another gives following equations (after dividing by 2):

$$a_1(P_1 + P_2) + 2a_{11}(P_1^3 + P_2^3) + a_{12}(P_1 P_2^2 + P_2 P_1^2) - \frac{g_{44}}{2}\frac{\partial^2(P_1+P_2)}{\partial x_3^2} = 0, \tag{A.2a}$$

$$a_1(P_1 - P_2) + 2a_{11}(P_1^3 - P_2^3) + a_{12}(P_1 P_2^2 - P_2 P_1^2) - \frac{g_{44}}{2}\frac{\partial^2(P_1-P_2)}{\partial x_3^2} = 0. \tag{A.2b}$$

For the case $a_{12} > -2a_{11}$, after introducing new order parameters, $P = P_1 + P_2$ and $A = P_1 - P_2$, one could get the following equations for $P$ and $A$:

$$a_1 P + \left(\frac{a_{11}}{2} + \frac{a_{12}}{4}\right) P^3 + \left(\frac{3}{2}a_{11} - \frac{a_{12}}{4}\right) P A^2 - \frac{g_{44}}{2}\frac{\partial^2 P}{\partial x_3^2} = 0, \tag{A.3a}$$

$$a_1 A + \left(\frac{a_{11}}{2} + \frac{a_{12}}{4}\right) A^3 + \left(\frac{3}{2}a_{11} - \frac{a_{12}}{4}\right) A P^2 - \frac{g_{44}}{2}\frac{\partial^2 A}{\partial x_3^2} = 0. \tag{A.3b}$$

The free energy density is $g_{LGD} = I_1 + \frac{g_{44}}{2}\left[\left(\frac{dP}{dx}\right)^2 + \left(\frac{dA}{dx}\right)^2\right]$, where first integral (that is an x-independent constant) is $I_1 = \frac{a_1}{2}(P^2 + A^2) + \left(\frac{a_{11}}{8} + \frac{a_{12}}{16}\right)(P^4 + A^4) + \left(\frac{3a_{11}}{4} - \frac{a_{12}}{8}\right) P^2 A^2 - \frac{g_{44}}{4}\left[\left(\frac{dP}{dx}\right)^2 + \left(\frac{dA}{dx}\right)^2\right]$.

Hence $g_{LGD} = \frac{a_1}{2}(P^2 + A^2) + \left(\frac{a_{11}}{8} + \frac{a_{12}}{16}\right)(P^4 + A^4) + \left(\frac{3a_{11}}{4} - \frac{a_{12}}{8}\right) P^2 A^2 + \frac{g_{44}}{4}\left[\left(\frac{dP}{dx}\right)^2 + \left(\frac{dA}{dx}\right)^2\right]$ which is similar to the expression via polarization components:

$$g_{LGD} = a_1(P_1^2 + P_2^2) + a_{11}(P_1^4 + P_2^4) + a_{12} P_1^2 P_2^2 + \frac{g_{44}}{2}\left[\left(\frac{dP_1}{dx}\right)^2 + \left(\frac{dP_2}{dx}\right)^2\right]$$

It is seen that Eqs.(A.3a) and (A.3b) are independent on each other only for the specific case $a_{12} = 6a_{11}$, and could be solved separately. At the simultaneous limit $m \to 1$ and $n \to 1$ it is easy to get a simpler expression from the solution (13):

$$P_1(x_3) = \frac{P_S}{2}\left[\tanh\left(\frac{x-x_0}{\sqrt{2}R_C}\right) + \tanh\left(\frac{x-x_a}{\sqrt{2}R_C}\right)\right] \equiv \frac{P_S}{2} \frac{\sinh\left(\frac{x-x_0}{\sqrt{2}R_C}\right)\cosh\left(\frac{x-x_a}{\sqrt{2}R_C}\right) + \cosh\left(\frac{x-x_0}{\sqrt{2}R_C}\right)\sinh\left(\frac{x-x_a}{\sqrt{2}R_C}\right)}{\cosh\left(\frac{x-x_0}{\sqrt{2}R_C}\right)\cosh\left(\frac{x-x_a}{\sqrt{2}R_C}\right)} \equiv$$

$$P_S \frac{\sinh\left(\frac{2x-x_0-x_a}{\sqrt{2}R_C}\right)}{\cosh\left(\frac{x_0-x_a}{\sqrt{2}R_C}\right) + \cosh\left(\frac{2x-x_0-x_a}{\sqrt{2}R_C}\right)} \quad (A.4a)$$

$$P_2(x_3) = \frac{P_S}{2}\left[\tanh\left(\frac{x-x_0}{\sqrt{2}R_C}\right) - \tanh\left(\frac{x-x_a}{\sqrt{2}R_C}\right)\right] \equiv P_S \frac{\sinh\left(\frac{x_a-x_0}{\sqrt{2}R_C}\right)}{\cosh\left(\frac{x_0-x_a}{\sqrt{2}R_C}\right) + \cosh\left(\frac{2x-x_0-x_a}{\sqrt{2}R_C}\right)} \quad (A.4b)$$

It is seen that Eqs.(A.4) coincides with equations (10) up to the change of designations.

### Substitution of the analytical solution (13) to the boundary conditions

**1) The case $\lambda_i = 0$.** The boundary conditions are

$$p_1(\pm l/2) = \frac{1}{2}\left[\sqrt{\frac{2m}{1+m}}sn\left(\frac{\pm l/2+x_{w1}}{\sqrt{1+m}}\bigg|m\right) + \sqrt{\frac{2n}{1+n}}sn\left(\frac{\pm l/2+x_{w2}}{\sqrt{1+n}}\bigg|n\right)\right] = 0, \quad (A.5a)$$

$$p_2(\pm l/2) = \frac{1}{2}\left[\sqrt{\frac{2m}{1+m}}sn\left(\frac{\pm l/2+x_{w1}}{\sqrt{1+m}}\bigg|m\right) - \sqrt{\frac{2n}{1+n}}sn\left(\frac{\pm l/2+x_{w2}}{\sqrt{1+n}}\bigg|n\right)\right] = 0, \quad (A.5b)$$

These Eqs.(A.5) are equivalent to the two sets of independent equations:

$$sn\left(\frac{-l/2+x_{w1}}{\sqrt{1+m}}\bigg|m\right) = sn\left(\frac{l/2+x_{w1}}{\sqrt{1+m}}\bigg|m\right) = 0, \quad (A.6a)$$

$$sn\left(\frac{-l/2+x_{w2}}{\sqrt{1+n}}\bigg|n\right) = sn\left(\frac{l/2+x_{w2}}{\sqrt{1+n}}\bigg|n\right) = 0, \quad (A.6b)$$

From these equations

$$\frac{-l/2+x_{w1}}{\sqrt{1+m}} = \pm 2\mathbf{K}(m)M_1, \qquad \frac{l/2+x_{w1}}{\sqrt{1+m}} = \pm 2\mathbf{K}(m)M_2. \quad (A.7a)$$

$$\frac{-l/2+x_{w2}}{\sqrt{1+n}} = \pm 2\mathbf{K}(m)N_1, \qquad \frac{l/2+x_{w2}}{\sqrt{1+n}} = \pm 2\mathbf{K}(m)N_2. \quad (A.7b)$$

Here $\mathbf{K}(y)$ is the complete elliptic integral of the first kind, which determines the half-period $2\mathbf{K}(y)$ of the elliptic functions. $M_i$ and $N_i$ are integers. Next, the summation and difference of Eqs.(A.7) gives that the modules, $m$ and $n$, satisfy the following transcendental equations:

$$l = \pm 2\sqrt{1+m}\mathbf{K}(m)(M_1 - M_2) \equiv 2\sqrt{1+m}\mathbf{K}(m)N_x, \quad (A.8a)$$

$$x_{w1} = \pm\sqrt{1+m}(M_1 + M_2)\mathbf{K}(m) = \pm\sqrt{1+m}(M_1 - M_2)\mathbf{K}(m) \pm 2M_2\mathbf{K}(m)\sqrt{1+m}, \quad (A.8b)$$

Since $sn(z + 2\mathbf{K}(m)M_2|m) \equiv (-1)^{M_2}sn(z|m)$, we obtained that $x_{w1} = \frac{l}{2}$ or $x_{w1} = \frac{l}{2} \pm 2\sqrt{1+m}\mathbf{K}(m)$. These two cases are $x_{w1} = \frac{l}{2} \pm (1-(-1)^k)\sqrt{1+m}\mathbf{K}(m)$, where $k$ is an integer. Similarly

$$l = 2\sqrt{1+n}\mathbf{K}(n)N_y, \quad (A.8c)$$

$$x_{w2} = \frac{l}{2}, \quad \text{or } x_{w2} = \frac{l}{2} \pm 2\sqrt{1+n}\mathbf{K}(n) \tag{A.8d}$$

Also we introduced the number of nodes, $N_x$ and $N_y$, which correspond to the solutions of Eqs.(11a) and (11b), respectively.

**1) The case $\lambda_i = +\infty$.** The boundary conditions are

$$\left.\frac{dp_1}{dx}\right|_{x=\pm l/2} = \frac{1}{2}\left[\frac{1}{\sqrt{1+m}}\sqrt{\frac{2m}{1+m}} cn\left(\frac{\pm l/2+x_{w1}}{\sqrt{1+m}}\middle| m\right) dn\left(\frac{\pm l/2+x_{w1}}{\sqrt{1+m}}\middle| m\right) + \right.$$

$$\left.\frac{1}{\sqrt{1+n}}\sqrt{\frac{2n}{1+n}} cn\left(\frac{\pm l/2+x_{w2}}{\sqrt{1+n}}\middle| n\right) dn\left(\frac{\pm l/2+x_{w2}}{\sqrt{1+n}}\middle| n\right)\right] = 0, \tag{A.9a}$$

$$\left.\frac{dp_2}{dx}\right|_{x=\pm l/2} = \frac{1}{2}\left[\frac{1}{\sqrt{1+m}}\sqrt{\frac{2m}{1+m}} cn\left(\frac{\pm l/2+x_{w1}}{\sqrt{1+m}}\middle| m\right) dn\left(\frac{\pm l/2+x_{w1}}{\sqrt{1+m}}\middle| m\right) - \right.$$

$$\left.\frac{1}{\sqrt{1+n}}\sqrt{\frac{2n}{1+n}} cn\left(\frac{\pm l/2+x_{w2}}{\sqrt{1+n}}\middle| n\right) dn\left(\frac{\pm l/2+x_{w2}}{\sqrt{1+n}}\middle| n\right)\right] = 0, \tag{A.9b}$$

These Eqs.(A.9) are equivalent to the two sets of independent equations:

$$cn\left(\frac{-l/2+x_{w1}}{\sqrt{1+m}}\middle| m\right) = cn\left(\frac{l/2+x_{w1}}{\sqrt{1+m}}\middle| m\right) = 0, \tag{A.10a}$$

$$cn\left(\frac{-l/2+x_{w2}}{\sqrt{1+n}}\middle| n\right) = cn\left(\frac{l/2+x_{w2}}{\sqrt{1+n}}\middle| n\right) = 0, \tag{A.10b}$$

From these equations

$$\frac{-l/2+x_{w1}}{\sqrt{1+m}} = \pm\mathbf{K}(m)(2M_1+1), \quad \frac{l/2+x_{w1}}{\sqrt{1+m}} = \pm\mathbf{K}(m)(2M_2+1). \tag{A.11a}$$

$$\frac{-l/2+x_{w2}}{\sqrt{1+n}} = \pm\mathbf{K}(m)(2N_1+1), \quad \frac{l/2+x_{w2}}{\sqrt{1+n}} = \pm\mathbf{K}(m)(2N_2+1). \tag{A.11b}$$

Here $\mathbf{K}(y)$ is the complete elliptic integral of the first kind, which determines the half-period $2\mathbf{K}(y)$ of the elliptic functions. $M_i$ and $N_i$ are integers. Next, the summation and difference of Eqs.(A.11) gives that the modules, $m$ and $n$, satisfy the following transcendental equations:

$$l = \pm 2\sqrt{1+m}\mathbf{K}(m)(M_1 - M_2) \equiv 2\sqrt{1+m}\mathbf{K}(m)N_x, \tag{A.12a}$$

$$x_{w1} = \pm\sqrt{1+m}(M_1+M_2+1)\mathbf{K}(m) = \frac{l}{2} \pm (2M_2+1)\mathbf{K}(m)\sqrt{1+m}. \tag{A.12b}$$

Since $cn(z + 2\mathbf{K}(m)M_2|m) \equiv (-1)^{M_2} cn(z|m)$

$$x_{w1} = \frac{l}{2} \pm \sqrt{1+m}\mathbf{K}(m), \quad x_{w2} = \frac{l}{2} \pm \sqrt{1+n}\mathbf{K}(n), \quad (\lambda_i \to +\infty). \tag{A.13}$$

**AIII. Dimensionless variables and the first integral**

To analyse the domain wall structure, we introduce the dimensionless coordinate x, polarization components $p_1$ and $p_1$, and ferroelectric anisotropy factor $\mu$:

$$x = \frac{x_3}{R_c}, \quad p_1 = \frac{P_1}{P_S}, \quad p_2 = \frac{P_2}{P_S}, \quad \mu = \frac{a_{12}}{2a_{11}}, \tag{A.5a}$$

Where we introduced the spontaneous polarization value $P_S = \sqrt{-a_1/(2a_{11})}$ and correlation length $R_c = \sqrt{-g_{44}/(2a_1)}$. Free energy density in dimensionless variables could be rewritten as

$$g_{LGD} = a_1(P_1^2 + P_2^2) + a_{11}(P_1^4 + P_2^4) + a_{12}P_1^2 P_2^2 + \frac{g_{44}}{2}\left[\left(\frac{dP_1}{dx_3}\right)^2 + \left(\frac{dP_2}{dx_3}\right)^2\right]$$

$$\equiv -\frac{a_1^2}{2a_{11}}(p_1^2 + p_2^2) + \frac{a_1^2}{4a_{11}}(p_1^4 + p_2^4) + \frac{a_1^2 a_{12}}{4a_{11}^2}p_1^2 p_2^2$$

$$+ \frac{g_{44}}{2}\left[\left(\frac{dp_1}{dx}\right)^2 + \left(\frac{dp_2}{dx}\right)^2\right]\left(\frac{-a_1}{2a_{11}}\right)\left(\frac{-2a_1}{g_{44}}\right)$$

$$\equiv \frac{a_1^2}{a_{11}}\left(-\frac{p_1^2 + p_2^2}{2} + \frac{p_1^4 + p_2^4}{4} + \frac{\mu}{2}p_1^2 p_2^2 + \frac{1}{2}\left[\left(\frac{dp_1}{dx}\right)^2 + \left(\frac{dp_2}{dx}\right)^2\right]\right)$$

Using new variables P and A and their dimensionless counterparts $p = P/P_S$, $a = A/P_S$,

$$\frac{a_{11}}{a_1^2}g_{LGD} == \frac{a_1}{2}\frac{a_{11}}{a_1^2}\left(\frac{-a_1}{2a_{11}}\right)(p^2 + a^2)$$

$$+ \left(\frac{-a_1}{2a_{11}}\right)^2 \frac{a_{11}}{a_1^2}\left(\left(\frac{a_{11}}{8} + \frac{a_{12}}{16}\right)(p^4 + a^4) + \left(\frac{3a_{11}}{4} - \frac{a_{12}}{8}\right)p^2 a^2\right)$$

$$+ \frac{a_{11}}{a_1^2}\frac{g_{44}}{4}\left(\frac{-a_1}{2a_{11}}\right)\left(\frac{-2a_1}{g_{44}}\right)\left[\left(\frac{dp}{dx}\right)^2 + \left(\frac{da}{dx}\right)^2\right]$$

$$\equiv -\frac{p^2 + a^2}{4} + \left(\frac{1+\mu}{4}\right)\frac{p^4 + a^4}{8} + \left(\frac{3-\mu}{4}\right)\frac{p^2 a^2}{4} + \frac{1}{4}\left[\left(\frac{dp}{dx}\right)^2 + \left(\frac{da}{dx}\right)^2\right]$$

Finally, one has the free energy as a functional of dimensionless order parameters

$$g_{LGD}\frac{a_{11}}{a_1^2} = -\frac{1}{2}(p_1^2 + p_2^2) + \frac{1}{4}(p_1^4 + p_2^4) + \frac{\mu}{2}p_1^2 p_2^2 + \frac{1}{2}\left[\left(\frac{dp_1}{dx}\right)^2 + \left(\frac{dp_2}{dx}\right)^2\right] \equiv \quad \text{(A.5a)}$$

$$-\frac{1}{4}(p^2 + a^2) + \frac{1}{8}\left(\frac{1+\mu}{4}\right)(p^4 + a^4) + \frac{1}{4}\left(\frac{3-\mu}{4}\right)p^2 a^2 + \frac{1}{4}\left[\left(\frac{dp}{dx}\right)^2 + \left(\frac{da}{dx}\right)^2\right] \quad \text{(A.5b)}$$

Variation of (A.5a) gives Euler-Lagrange equations

$$\frac{\partial^2}{\partial x^2}p_1 = -p_1 + p_1^3 + \mu p_1 p_2^2, \quad \text{(A.6a)}$$

$$\frac{\partial^2}{\partial x^2}p_2 = -p_2 + p_2^3 + \mu p_2 p_1^2, \quad \text{(A.6b)}$$

After elementary transformations of Eq.(A.6), namely $\frac{\partial}{\partial x}p_1$ (A.6a) $+ \frac{\partial}{\partial x}p_2$ (A.6b), we obtained that the first integral exists:

$$I_1[\mu] = -\frac{1}{2}(p_1^2 + p_2^2) + \frac{1}{4}(p_1^4 + p_2^4) + \frac{\mu}{2}p_1^2 p_2^2 - \frac{1}{2}\left[\left(\frac{dp_1}{dx}\right)^2 + \left(\frac{dp_2}{dx}\right)^2\right] \quad \text{(A.7)}$$

Using the transformation, $p_1$ (A.6a) $+ p_2$ (A.6b), we obtained that

$$p_1 \frac{\partial^2 p_1}{\partial x^2} + p_2 \frac{\partial^2 p_2}{\partial x^2} = -(p_1^2 + p_2^2) + (p_1^4 + p_2^4) + 2\mu p_1^2 p_2^2, \quad \text{(A.8a)}$$

Account for the identities, $\frac{\partial^2}{\partial x^2}p_1^2 = 2p_1 \frac{\partial^2 p_1}{\partial x^2} + 2\left(\frac{dp_1}{dx}\right)^2$, and $\frac{\partial^2}{\partial x^2}p_2^2 = 2p_2 \frac{\partial^2 p_2}{\partial x^2} + 2\left(\frac{dp_2}{dx}\right)^2$, in Eq.(A.8a), we obtain that

$$p_1 \frac{\partial^2 p_1}{\partial x^2} + p_2 \frac{\partial^2 p_2}{\partial x^2} = \frac{1}{2}\frac{\partial^2}{\partial x^2}(p_1^2 + p_2^2) - \left[\left(\frac{dp_1}{dx}\right)^2 + \left(\frac{dp_2}{dx}\right)^2\right] = -(p_1^2 + p_2^2) + (p_1^4 + p_2^4) + 2\mu p_1^2 p_2^2,$$

(A.8b)

$$\left[\left(\frac{dp_1}{dx}\right)^2 + \left(\frac{dp_2}{dx}\right)^2\right] = \frac{1}{2}\frac{\partial^2}{\partial x^2}(p_1^2 + p_2^2) + (p_1^2 + p_2^2) - (p_1^4 + p_2^4) - 2\mu p_1^2 p_2^2, \quad \text{(A.8c)}$$

From Eq. (A.8c)

$$I_1[\mu] = -\frac{1}{4}\frac{\partial^2}{\partial x^2}(p_1^2 + p_2^2) - (p_1^2 + p_2^2) + \frac{3}{4}(p_1^2 + p_2^2)^2 + \frac{3}{2}(\mu - 1)p_1^2 p_2^2 \quad \text{(A.9a)}$$

$$g_{LGD} = \frac{1}{4}\frac{\partial^2}{\partial x^2}(p_1^2 + p_2^2) - \frac{1}{4}(p_1^2 + p_2^2)^2 - \frac{\mu - 1}{2}p_1^2 p_2^2$$

$$= -(p_1^2 + p_2^2) + \frac{1}{2}(p_1^2 + p_2^2)^2 + (\mu - 1)p_1^2 p_2^2 - I_1[\mu]. \quad \text{(A.9b)}$$

Introducing new functions:

$$q = (p_1^2 + p_2^2), \quad s = p_1^2 p_2^2 \quad \text{(A.10a)}$$

$$(p_1 + p_2)^2 = q + 2\sqrt{s}, \quad (p_1 - p_2)^2 = q - 2\sqrt{s} \quad \text{(A.10b)}$$

We obtained that the first integral and free energy density become:

$$I_1[\mu] = -\frac{1}{4}\frac{\partial^2}{\partial x^2}q - q + \frac{3}{4}q^2 + \frac{3}{2}(\mu - 1)s, \quad \text{(A.11a)}$$

$$g_{LGD}[q,s] = \frac{1}{4}\frac{\partial^2}{\partial x^2}q - \frac{1}{4}q^2 - \frac{\mu - 1}{2}s. \quad \text{(A.11b)}$$

Two homogeneous phases are consistent with the energy (A.5):

$$p_1 = p_2 = \pm\frac{1}{\sqrt{1+\mu}}, \quad g_{LGD} = -\frac{1}{2(1+\mu)}, \quad I_1[\mu] = -\frac{1}{2(1+\mu)} \quad \text{stable at } -1 < \mu < 1, \quad \text{(A.12a)}$$

$$p_1^2 = 0, \; p_2^2 = 1, \text{ or } p_1^2 = 1, \; p_2^2 = 0, \; g_{LGD} = -\frac{1}{4}, \; I_1[\mu] = -\frac{1}{4} \text{ stable at } \mu > 1, \quad \text{(A.12b)}$$

Energy calculations show that the phase (A.12a) is stable at $-1 < \mu < 1$, while the paraelectric phase is always unstable below Curie temperature. The special case is $\mu = 1$.

Direct variational method can be applied for minimization Eq.(A.5) using the trial functions:

$$p_1(x) = p_{a1} \tanh\left(\frac{x+x_w}{w}\right) + p_{b1} \tanh\left(\frac{x-x_w}{w}\right), \quad \text{(A.13a)}$$

$$p_2(x) = p_{a2} \tanh\left(\frac{x+x_w}{w}\right) - p_{b2} \tanh\left(\frac{x-x_w}{w}\right) + c\left[1 - p_c \cosh^{-2}\left(\frac{x}{w}\right)\right], \quad \text{(A.13b)}$$

We obtained that $p_{a1} \approx p_{b1}$, $p_{a2} \approx p_{b2}$ for all $-1 < \mu < 10$. Also we obtained that $p_c = 0$ for all $-1 < \mu < 1$ and $\mu > 1.1$, except for the immediate vicinity of the special case, $\mu = 1$, when the numerical algorithm is unstable.

## APPENDIX B. Direct variational method

Let us apply the direct variational method for the minimization of the free energy functional with density given by Eqs.(15):

$$G = \int_{-\frac{L}{2}}^{\frac{L}{2}} dx \left(-\frac{1}{2}(p_1^2 + p_2^2) + \frac{1}{4}(p_1^4 + p_2^4) + \frac{\mu}{2}p_1^2 p_2^2 + \frac{1}{2}\left[\left(\frac{dp_1}{dx}\right)^2 + \left(\frac{dp_2}{dx}\right)^2\right]\right)\Bigg|_{L \to \infty} \quad \text{(B.1)}$$

using the trial functions:

$$p_1(x) = \frac{1}{2} a_1 \left[\tanh\left(\frac{x}{w} + \frac{\xi}{2}\right) + \tanh\left(\frac{x}{w} - \frac{\xi}{2}\right)\right], \quad (B.2a)$$

$$p_2(x) = a_2 + \frac{b_2}{\xi}\left[\tanh\left(\frac{x}{w} + \frac{\xi}{2}\right) - \tanh\left(\frac{x}{w} - \frac{\xi}{2}\right)\right], \quad (B.2b)$$

where $a_1$, $a_2$, $\xi$, $w$, and $b_2$ are variational parameters, which can be determined after substitution of Eqs.(B.2) in the free energy (B.1), further integration and minimization of the integral $\int_{-L}^{L} g_{LGD}(x) dx$ over these parameters. Note that we used new designation $\xi = 2x_w/w$.

To verify how accurate are the trial functions (B.2), the deviation of the first integral (A.5a) from the constant value, $I_1[\mu] = -\frac{a_1^2}{2} + \frac{a_1^4}{4}$, obtained for $L \to \infty$, can be estimated. At the domain wall plane $x = 0$, the trial functions (B.2) gives for the first integral $I_1[\mu] = -\frac{1}{2}\left(2\frac{b_2}{\xi}\tanh\left(\frac{\xi}{2}\right) + a_2\right)^2 + \frac{1}{4}\left(2\frac{b_2}{\xi}\tanh\left(\frac{\xi}{2}\right) + a_2\right)^4 - \frac{a_1^2}{2w^2}\left(\cosh\left(\frac{\xi}{2}\right)\right)^{-4}$.

This way, in the case of analytical expressions for the integrals, allows us to obtain analytical dependencies for the variational parameters on the master parameter $\mu$. The results of the integration could be written as follows:

$$G = G_L + G_{DW} \quad (B.3a)$$

Where Landau energy $G_L$ is proportional to the system size and independent on the domain wall parameters:

$$G_L = L\left(-\frac{a_1^2 + a_2^2}{2} + \frac{a_1^4 + a_2^4}{4} + \frac{\mu}{2} a_1^2 a_2^2\right). \quad (B.3b)$$

Other integrands are independent on the system size $L$:

$$G_{DW} = \frac{a_1^2}{w} B_1(\xi) + \frac{b_2^2}{w} B_2(\xi) + w A_1(\xi) a_1^2 + w A_2(\xi) b_2^2 - 2 w a_2 b_2 + \mu w a_1^2 a_2 b_2 M_{12}^{(3)}(\xi) +$$
$$+ \mu w a_1^2 b_2^2 M_{12}^{(4)}(\xi) + \mu w a_1^2 a_2^2 M_1^{(2)}(\xi) + 2 w a_2^3 b_2 + w b_2^2 a_2^2 F_2^{(2)}(\xi) + w b_2^3 a_2 F_2^{(3)}(\xi) +$$
$$+ w F_1^{(4)}(\xi) a_1^4 + w F_2^{(4)}(\xi) b_2^4 \quad (B.3c)$$

Here we used the following designations for the expansion coefficients:

$$A_1(\xi) = \frac{1 + \xi \coth(\xi)}{2} = \begin{cases} 1 + \frac{\xi^2}{6} & \text{at } \xi \to 0; \\ \frac{1+\xi}{2} & \text{at } \xi \to \infty. \end{cases} \quad (B.4a)$$

$$A_2(\xi) = \frac{2}{\xi^2}[1 - \xi \coth(\xi)] = \begin{cases} -\frac{2}{3} & \text{at } \xi \to 0; \\ \frac{2-2\xi}{\xi^2} & \text{at } \xi \to \infty. \end{cases} \quad (B.4b)$$

$$B_1(\xi) = \frac{1}{3} - \frac{1 - \xi \coth(\xi)}{\sinh(\xi)^2} = \begin{cases} \frac{2}{3} - \frac{2\xi^2}{15} & \text{at } \xi \to 0; \\ \frac{1}{3} + 4\xi \exp(-2\xi) & \text{at } \xi \to \infty. \end{cases} \quad (B.4c)$$

$$B_2(\xi) = \frac{4}{3\xi^2}\left(1 + 3\frac{1-\xi\coth(\xi)}{\sinh(\xi)^2}\right) = \begin{cases} \frac{8}{15} & \text{at } \xi \to 0; \\ \frac{4}{3\xi^2} - \frac{16}{\xi}\exp(-2\xi) & \text{at } \xi \to \infty. \end{cases} \quad (B.4d)$$

$$M_{12}^{(3)}(\xi) = \frac{\sinh(2\xi)-2\xi}{2\xi\sinh(\xi)^2} = \begin{cases} \frac{2}{3} - \frac{4\xi^2}{45} & \text{at } \xi \to 0; \\ \frac{1}{\xi} - 4\exp(-2\xi) & \text{at } \xi \to \infty. \end{cases} \quad (B.4e)$$

$$M_{12}^{(4)}(\xi) = \frac{5+\cosh(2\xi)}{6\xi^2\sinh(\xi)^2} - \frac{\coth(\xi)}{\xi\sinh(\xi)^2} = \begin{cases} \frac{4}{15} & \text{at } \xi \to 0 \\ \frac{1}{3\xi^2} - \frac{4}{\xi}\exp(-2\xi) & \text{at } \xi \to \infty \end{cases} \quad (B.4f)$$

$$M_1^{(2)}(\xi) = -\frac{1}{2}\left[1 + \frac{\xi}{2}\coth(\xi)\right] = \begin{cases} -\frac{3}{4} - \frac{\xi^2}{12} & \text{at } \xi \to 0; \\ -\frac{1}{2} - \frac{\xi}{4} & \text{at } \xi \to \infty. \end{cases} \quad (B.4g)$$

$$F_2^{(2)}(\xi) = \frac{6}{\xi^2}[-1 + \xi\coth(\xi)] = \begin{cases} 2 & \text{at } \xi \to 0; \\ \frac{6}{\xi^2}(\xi-1) & \text{at } \xi \to \infty. \end{cases} \quad (B.4h)$$

$$F_2^{(3)}(\xi) = 2\frac{2\xi[2+\cosh(2\xi)]-3\sinh(2\xi)}{\xi^3\sinh(\xi)^2} = \begin{cases} \frac{16}{15} & \text{at } \xi \to 0; \\ \frac{4(2\xi-3)}{\xi^3} & \text{at } \xi \to \infty. \end{cases} \quad (B.4i)$$

$$F_1^{(4)}(\xi) = -\frac{11\sinh(\xi)^2 + 3 + 3\xi\coth(\xi)[2\sinh(\xi)^2-1]}{24\sinh(\xi)^2} = \begin{cases} -\frac{2}{3} - \frac{\xi^2}{10} & \text{at } \xi \to 0; \\ \frac{-6\xi-11}{24} & \text{at } \xi \to \infty. \end{cases} \quad (B.4j)$$

$$F_2^{(4)}(\xi) = \frac{6\xi[9\cosh(\xi)+\cosh(3\xi)]-27\sinh(\xi)-11\sinh(3\xi)}{6\xi^4\sinh(\xi)^3} = \begin{cases} \frac{8}{35} & \text{at } \xi \to 0; \\ \frac{2(6\xi-11)}{3\xi^4} & \text{at } \xi \to \infty. \end{cases} \quad (B.4k)$$

All of them is the functions of parameter $\xi$ only. Note the first equalities are the exact integrals, while the second sets are the limiting cases for the small ($\xi \ll 1$) and high ($\xi \gg 1$) value of parameter $\xi$.

After the integration the free energy (B.3a) becomes an algebraic function of $a_1$, $a_2$, $w$, $b_2$, and a transcendental function of $\xi$. The first step is the minimization of $G_L$ in Eq. (B.3b) with respect to $a_1$ and $a_2$, which gives the following equations:

$$-a_1 + a_1^3 + \mu\, a_1 a_2^2 = 0 \quad (B.5a)$$
$$-a_2 + a_2^3 + \mu\, a_1^2 a_2 = 0 \quad (B.5b)$$

The system (B.5) have several solutions, namely:

I. Paraelectric phase "P" corresponding to a trivial solution for both order parameters, $a_1 = 0$ and $a_2 = 0$. P-phase is not realized entire $\mu$-range in the considered case, since the parameters are well below Curie temperature.

II. "Tetragonal" polar phase "T₁" corresponding to $a_1 = \pm 1$ and $a_2 = 0$. T₁-phase is realized for $\mu > 1$

III. "Tetragonal" polar phase "T$_2$" corresponding to $a_1 = 0$ and $a_2 = \pm 1$. T$_2$-phase is not realized entire $\mu$-range

IV. "Orthogonal" polar phase "O" corresponding to $a_1 = \pm\sqrt{\frac{1}{1+\mu}}$ and $a_2 = \pm\sqrt{\frac{1}{1+\mu}}$, where the signs are independent. O-phase is realized for $-1 < \mu < 1$.

Further consideration is based on the minimization of Eq.(B.3c) with respect to the variational parameters $b_2, \xi$ and $w$ after substituting the evident expressions for $a_1$ and $a_2$.

I. "P" phase – no domain walls, with $a_1, a_2, w$ and $b_2$ equal to zero.

II. In "T$_1$" phase (with $a_1 = \pm 1$ and $a_2 = 0$) the domain wall contribution (B.3c) to the free energy could simplified as follows:

$$G_{DW} = \frac{1}{w}\left(B_1(\xi) + b_2^2 B_2(\xi)\right) + w\left(A_1(\xi) + A_2(\xi) b_2^2 + \mu b_2^2 M_{12}^{(4)}(\xi) + F_1^{(4)}(\xi) + F_2^{(4)}(\xi) b_2^4\right)$$
(B.6)

and minimization with respect to the variational parameters $b_2, \xi$ and $w$ gives the following equations:

$$\frac{\partial G_{DW}}{\partial w} = -\frac{B_1(\xi) + b_2^2 B_2(\xi)}{w^2} + \left(A_1(\xi) + A_2(\xi) b_2^2 + \mu b_2^2 M_{12}^{(4)}(\xi) + F_1^{(4)}(\xi) + F_2^{(4)}(\xi) b_2^4\right) = 0 \quad \text{(B.7a)}$$

$$\frac{\partial G_{DW}}{\partial b_2} = \frac{2 b_2 B_2(\xi)}{w} + 2 b_2 w\left(A_2(\xi) + \mu M_{12}^{(4)}(\xi) + 2 F_2^{(4)}(\xi) b_2^2\right) = 0 \quad \text{(B.7b)}$$

$$\frac{\partial G_{DW}}{\partial \xi} = 0 \quad \Rightarrow$$

$$\frac{1}{w}\left(\frac{\partial B_1(\xi)}{\partial \xi} + b_2^2 \frac{\partial B_2(\xi)}{\partial \xi}\right) + w\left(\frac{\partial A_1(\xi)}{\partial \xi} + \frac{\partial A_2(\xi)}{\partial \xi} b_2^2 + \mu b_2^2 \frac{\partial M_{12}^{(4)}(\xi)}{\partial \xi} + \frac{\partial F_1^{(4)}(\xi)}{\partial \xi} + \frac{\partial F_2^{(4)}(\xi)}{\partial \xi} b_2^4\right) = 0$$
(B.7c)

III. In "T$_2$" phase (with $a_1 = 0$ and $a_2 = \pm 1$) the domain wall contribution (B.3c) to the free energy could essentially simplified as follows:

$$G_{DW} = \frac{b_2^2}{w} B_2(\xi) + w\left(A_2(\xi) b_2^2 + b_2^2 F_2^{(2)}(\xi) + b_2^3 a_2 F_2^{(3)}(\xi) + F_2^{(4)}(\xi) b_2^4\right) \quad \text{(B.8)}$$

where $a_2 = \pm 1$. Since always $w b_2^2 F_2^{(3)}(\xi) > 0$, minimization in Eq.(B.7a) leads to condition $a_2 b_2 < 0$ for a given $a_2$-sign.

IV. In "O" phase the domain wall contribution (B.3c) to the free energy could rewritten as follows:

$$G_{DW} = \frac{1}{w(1+\mu)} B_1(\xi) + \frac{b_2^2}{w} B_2(\xi) + w\left(\frac{A_1(\xi)}{1+\mu} + \frac{F_1^{(4)}(\xi)}{(1+\mu)^2} + \frac{\mu}{(1+\mu)^2} M_1^{(2)}(\xi) + \frac{\mu}{1+\mu} a_2 b_2 M_{12}^{(3)}(\xi) -\right.$$

$$\left. 2 a_2 b_2 \frac{\mu}{1+\mu} + A_2(\xi) b_2^2 + \frac{\mu}{1+\mu} b_2^2 M_{12}^{(4)}(\xi) + b_2^2 \frac{F_2^{(2)}(\xi)}{1+\mu} + b_2^3 a_2 F_2^{(3)}(\xi) + F_2^{(4)}(\xi) b_2^4\right), \quad \text{(B.9a)}$$

Here we used that $a_2 = \pm\sqrt{\frac{1}{1+\mu}}$.

Let us consider the case when $\xi \to 0$, here we obtained from Eq.(B.9a) the following expression for the energy:

$$G_{DW} = \frac{2}{3w(1+\mu)} + \frac{8b_2^2}{15w} + w\left(-\frac{4}{3}\frac{1}{1+\mu} - \frac{3}{4}\frac{\mu}{(1+\mu)^2} - \frac{4}{3}\frac{\mu}{1+\mu}a_2b_2 + b_2^2\frac{20-6\mu}{15(1+\mu)} + \frac{16}{15}b_2^3a_2 + \frac{8}{35}b_2^4\right)$$

(B.9b)

Its minimization with respect to the variational parameters $b_2$ and $w$ gives the following equations:

$$\frac{\partial G_{DW}}{\partial w} = -\frac{\left(\frac{2}{3(1+\mu)} + \frac{8b_2^2}{15}\right)}{w^2} + \left(-\frac{4}{3}\frac{1}{1+\mu} - \frac{3}{4}\frac{\mu}{(1+\mu)^2} - \frac{4}{3}\frac{\mu}{1+\mu}a_2b_2 + b_2^2\frac{20-6\mu}{15(1+\mu)} + \frac{16}{15}b_2^3a_2 + \frac{8}{35}b_2^4\right) = 0$$

(B.10a)

$$\frac{\partial G_{DW}}{\partial b_2} = +\frac{16}{15w}b_2 + w\left(-\frac{4}{3}\frac{\mu}{1+\mu}a_2 + 2\frac{20-6\mu}{15(1+\mu)}b_2 + \frac{48}{15}b_2^2a_2 + \frac{32}{35}b_2^3\right) = 0 \quad \text{(B.10b)}$$

Eq.(B.10b) could be rewritten as

$$\frac{1}{w^2} = -\frac{15}{16}\left(-\frac{4}{3}\frac{\mu}{1+\mu}\frac{a_2}{b_2} + 2\frac{20-6\mu}{15(1+\mu)} + \frac{48}{15}b_2a_2 + \frac{32}{35}b_2^2\right) \quad \text{(B.11a)}$$

The substitution of (B.11a) into the Eq.(B.10a) gives

$$\left(\frac{5}{4(1+\mu)} + b_2^2\right)\left(-\frac{2}{3}\frac{\mu}{1+\mu}\frac{a_2}{b_2} + \frac{20-6\mu}{15(1+\mu)} + \frac{24}{15}b_2a_2 + \frac{16}{35}b_2^2\right) + \left(-\frac{4}{3}\frac{1}{1+\mu} - \frac{3}{4}\frac{\mu}{(1+\mu)^2} - \frac{4}{3}\frac{\mu}{1+\mu}a_2b_2 + b_2^2\frac{20-6\mu}{15(1+\mu)} + \frac{16}{15}b_2^3a_2 + \frac{8}{35}b_2^4\right) = 0 \quad \text{(B.11b)}$$

Using supposition $w^2 \approx 2$ (valid near zero point of μ) one could get from (B.11a) the following

$$\frac{12-\mu}{4(1+\mu)} \approx \frac{5}{4}\frac{\mu}{1+\mu}\frac{a_2}{b_2} \Rightarrow b_2 \approx \frac{5}{12}\mu \quad \text{(B.12)}$$

Near the critical point $\mu \to -1$ one has $b_2 = -\eta/\sqrt{1+\mu}$ (with η independent on μ), while $a_2 = \sqrt{\frac{1}{1+\mu}}$, which gives the algebraic equation for η

$$\left(\frac{5}{4} + \eta^2\right)\left(-\frac{2}{3\eta} + \frac{26}{15} - \eta\frac{24}{15} + \eta^2\frac{16}{35}\right) + \left(\frac{3}{4} - \frac{4}{3}\eta + \eta^2\frac{26}{15} - \frac{16}{15}\eta^3 + \eta^4\frac{8}{35}\right) \approx 0 \quad \text{(B.13)}$$

It could be easily shown that η≈0.52